\documentclass[
reprint,
 aps,
 superscriptaddress,
bibnotes,
pra
]{revtex4-2}
\usepackage{physics}
\usepackage{dsfont}
\usepackage{color,newfloat}
\usepackage{graphicx}
\usepackage{dcolumn}
\usepackage{bm}
\usepackage{amsmath,amssymb,amsthm}
\usepackage{mathtools}
\usepackage{multirow}
\usepackage{hhline}
\usepackage{comment}
\usepackage[dvipsnames]{xcolor}
\usepackage{hyperref}
\hypersetup{
    colorlinks=true,
    linkcolor=Maroon,
    citecolor=Maroon,
    urlcolor=Maroon
}
\usepackage{gensymb} % for \degree degree symbol

\usepackage[caption = false]{subfig}
\usepackage{lineno}
\usepackage{siunitx}
\usepackage[mathscr]{euscript}
 \let\mathscr\relax

\usepackage{makecell}

\begin{document}

%%%%%%%%%%%%%%%%%%%%%%%%%%%%%%%%%%%%%%%%%%%%
%%%%%%%%    Title/Authors/Abstract    %%%%%%
%%%%%%%%%%%%%%%%%%%%%%%%%%%%%%%%%%%%%%%%%%%%

\title{
Electrical control of quantum dots in GaAs-on-insulator waveguides for coherent single-photon generation}

\author{Hanna Salamon}
\email{hanna.salamon@nbi.ku.dk}
\affiliation{Center for Hybrid Quantum Networks (Hy-Q), Niels Bohr Institute, University of Copenhagen, Jagtvej 155A, 2200 Copenhagen, Denmark}

\author{Ying Wang}
\email{ying.wang@nbi.ku.dk}
\affiliation{Center for Hybrid Quantum Networks (Hy-Q), Niels Bohr Institute, University of Copenhagen, Jagtvej 155A, 2200 Copenhagen, Denmark}

\author{Arnulf Snedker-Nielsen}
\affiliation{Center for Hybrid Quantum Networks (Hy-Q), Niels Bohr Institute, University of Copenhagen, Jagtvej 155A, 2200 Copenhagen, Denmark}

\author{Atefeh Shadmani}
\affiliation{Center for Hybrid Quantum Networks (Hy-Q), Niels Bohr Institute, University of Copenhagen, Jagtvej 155A, 2200 Copenhagen, Denmark}

\author{R{\"u}diger Schott}
\affiliation{Lehrstuhl f{\"u}r Angewandte Festk{\"o}rperphysik, Ruhr-Universit{\"a}t Bochum, Universit{\"a}tsstrasse 150, D-44780 Bochum, Germany}

\author{Mircea Balauroiu}
\affiliation{Department of Electrical and Computer Engineering, Aarhus University, 8200 Aarhus N, Denmark}
\author{Nicolas Volet}
\affiliation{Department of Electrical and Computer Engineering, Aarhus University, 8200 Aarhus N, Denmark}

\author{Arne Ludwig}
\affiliation{Lehrstuhl f{\"u}r Angewandte Festk{\"o}rperphysik, Ruhr-Universit{\"a}t Bochum, Universit{\"a}tsstrasse 150, D-44780 Bochum, Germany}

\author{Leonardo~Midolo}
\affiliation{Center for Hybrid Quantum Networks (Hy-Q), Niels Bohr Institute, University of Copenhagen, Jagtvej 155A, 2200 Copenhagen, Denmark}

%%%%%%%%%%%%%%%%%%%%%%%%%%%%%%%%%%
%%%%%%%%     Abstract    %%%%%%%%%
%%%%%%%%%%%%%%%%%%%%%%%%%%%%%%%%%%
\begin{abstract}
The integration of coherent quantum emitters with silicon photonic platforms is essential for scalable quantum technologies. We demonstrate electrically controlled self-assembled quantum dots embedded in GaAs waveguides bonded onto a SiO$_2$/Si substrate and coupled to low-loss SiN waveguides. Our approach uses a die-to-die adhesive bonding process to realize a GaAs-on-insulator platform incorporating a \emph{p-i-n} junction for charge noise suppression and Stark tuning of excitonic transitions. Resonance fluorescence measurements reveal narrow optical linewidths below 2 $\mu$eV and high single-photon purity, with ${g^{(2)}(0) = (5.2 \pm 0.8)\%}$, matching the performance of unprocessed GaAs devices. These results establish a practical route to integrate high-coherence quantum light sources with mature silicon photonics, enabling scalable quantum photonic integrated circuits. 
\end{abstract}

\date{\today}

\maketitle

%%%%%%%%%%%%%%%%%%%%%%%%%%%%%%%%%%%
%%%%%%%%     Main text    %%%%%%%%%
%%%%%%%%%%%%%%%%%%%%%%%%%%%%%%%%%%%
Building photonic integrated circuits (PICs) with solid-state quantum emitters is a crucial requirement to advance optical quantum networks and future quantum technologies. Among available emitters, self-assembled quantum dots (QDs) in gallium arsenide (GaAs) stand out because they combine high radiative efficiency, low noise \cite{pinotsi2011resonant, lobl2017narrow, zhai2020low}, and near-transform-limited linewidths \cite{kuhlmann2015transform}, making them ideal two-level systems for quantum information science \cite{lodahl2015interfacing, arakawa2020progress, heindel2023quantum} and near-deterministic single-photon generation \cite{uppu2020scalable, tomm2021bright, liu2018high}. 
From the perspective of photonic integration, however, GaAs is still lagging behind compared to silicon photonics or other III/V materials, due to the higher optical loss and lack of well-established manufacturing methods, PIC suppliers, and foundry services. Unlike established PIC platforms, such as silicon-on-insulator (SOI), thin-film lithium niobate (TFLN), and silicon nitride (SiN), epitaxially-grown GaAs lacks a low-index substrate or cladding material, which is necessary to achieve near-unity emitter-photon coupling in single-mode waveguides \cite{arcari2014near, kirvsanske2017indistinguishable, morais2017directionally}. A practical solution is to heterogeneously integrate GaAs with low-index materials like silica (SiO$_{2}$), forming GaAs-on-insulator waveguides with high refractive-index contrast and low propagation loss \cite{zhou2024gaas, chang2018heterogeneously, stanton2020efficient}. Using silica as low-index substrate offers matched thermal expansion with GaAs, enhances the mechanical stability at cryogenic temperatures, and minimizes the risk of structural collapse often observed in fully-suspended GaAs waveguides. Additionally, SiO$_{2}$ is compatible with complementary metal-oxide-semiconductor (CMOS) processing and is widely used as a cladding material in platforms such as SiN and TFLN \cite{rigal2017propagation, wang2021electroabsorption, pfister2025telecom, chanana2022ultra, descamps2024acoustic}.
Several strategies have been explored to integrate III-V quantum emitters with silicon photonics including micro-transfer printing \cite{katsumi2022cmos, liang2010hybrid, osada2019strongly}, pick-and-place \cite{kim2017hybrid, chandrasekar2022mechanophotonics, chanana2022ultra, larocque2024tunable}, and die-to-die bonding \cite{shadmani2022integration, davanco2017heterogeneous, schnauber2019indistinguishable}. However, no prior work has demonstrated coherent photon generation from heterogeneously integrated emitters, an essential requirement for quantum technologies. To fully leverage heterogeneous integration and its scalability, the emitters should not only couple efficiently to a low-loss circuit, but also maintain low noise and minimal blinking which are critical for applications such as photon-mediated nonlinearities \cite{le2022dynamical} and entangled-photon generation \cite{meng2024deterministic}. 

In this work, we investigate whether high coherence and narrow optical linewidths can be preserved after transferring QDs from their native GaAs substrate to a different platform. We demonstrate a practical and inexpensive die-to-die bonding process that integrates GaAs membranes with embedded \emph{p-i-n} diodes onto a $SiO_2$ platform. Finally, we show that this approach is compatible with low-loss SiN PICs fabricated in a commercial foundry, providing a pathway towards scalable Si-based quantum photonics integration.

\begin{figure*}[h!tbp]
    \centering
    \includegraphics[width=\textwidth]{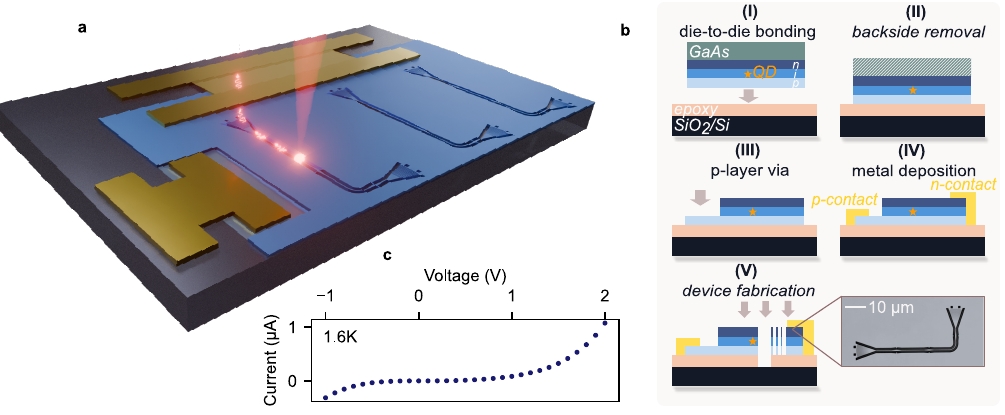}
    \caption{(a) Schematic of the GaAs-on-insulator platform featuring a  GaAs membrane (blue) with embedded quantum dots (QDs) and a \emph{p-i-n} junction bonded to a SiO$_2$ substrate (black). The QDs in single-mode nanobeam waveguides are optically excited and coupled off-chip via grating couplers.
    (b) Outline of the nanofabrication process (see text for details). The inset shows an optical microscope image of the nanobeam waveguide.
    (c) Current-voltage characteristic of the \emph{p-i-n} Schottky diode at cryogenic temperature (1.6 K) with low leakage current in the operation range of the device of 1-1.5 V.
    }
    
\label{fig:1}
\end{figure*} 
\section{GaAs-on-insulator waveguides}

Figure \ref{fig:1}a illustrates the concept of the hybrid GaAs-on-insulator platform, as a stepping stone towards integration with several Si-based waveguide platforms. The key element to achieve low-noise operation is the development of electrical contacts on the GaAs membranes to screen QDs from charge noise and enable Stark tuning \cite{grim2019scalable, patel2010two}. The QDs are grown on GaAs substrate using molecular beam epitaxy (MBE) and embedded within a thin \emph{p-i-n} diode structure. To form a baseline of comparison, we employ the same wafer and QD material characterized in detail in Ref. \cite{lobl2017narrow}. This configuration allows precise control of the local charge environment and yields narrow optical linewidths as previously reported for unprocessed samples.
The fabrication process is illustrated in Fig.~\ref{fig:1}(b) (details are provided in the Supporting Information). We employ adhesive die-to-die bonding (I) with an epoxy-based negative tone photoresist spin-coated on a target silica chip. The backside is removed (II) via a selective chemical etching, resulting in the whole 170-nm-thick GaAs membrane with embedded diode junction transferred on SiO$_2$. Openings to the buried $p-$layer (III) are made via reactive ion etching, followed by metal deposition (IV) of $p-$ and $n-$ contacts. Single-mode nanobeam waveguides and shallow-etched focusing grating couplers are patterned and etched in the last step (V).  The inset of Fig.~\ref{fig:1}(b) shows an optical microscope image of the fabricated GaAs-on-insulator waveguide with embedded QDs. 
The electrical operation of the diode was verified at cryogenic temperatures as shown in the current-voltage measurement in Fig.~\ref{fig:1}(c). The diode is designed to operate in forward bias, up to 1.5 V, to partially cancel the built-in electric field of the diode and control the charge state of the emitter. Low leakage currents $< 0.1 \mu$A are observed, which could be further suppressed by surface passivation. Importantly, both the voltage levels and electrical power consumption requirements ($< 100$ nW) to stabilize and tune QDs are well within the range of CMOS electronics, offering a great potential for future scalability in opto-electronic circuits.  

\section{Resonance fluorescence from quantum dots}

\begin{figure}
    \centering
    \includegraphics{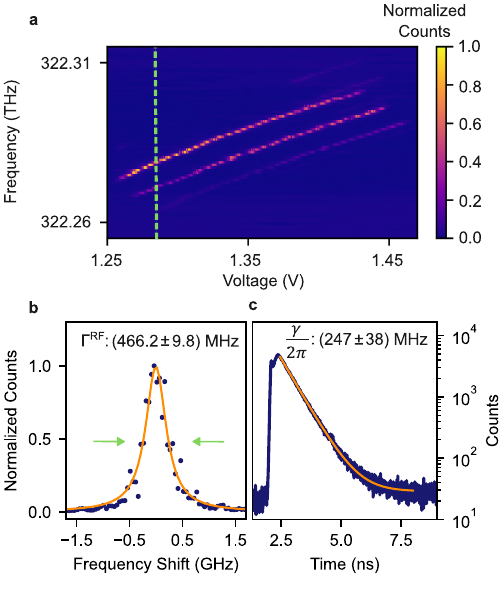}
    \caption{
    (a) Resonance fluorescence from quantum dots excited by a continuous-wavelength (CW) laser as a function of laser frequency and bias voltage. 
    (b) Scan of a single QD transition (wavelength of 930.6 nm) indicated by the green dashed line in (a). A Voigt profile fit (solid orange line) yields a full width at half maximum (FWHM) of $\Gamma^{RF} = (466.2 \pm 9.8)$ MHz.
    (c) Time-resolved fluorescence with a pulsed laser at $\pi$-pulse area, showing a single-exponential decay rate $\gamma = (1.52 \pm 0.01)$ ns$^{-1}$.
    }
    \label{fig:2}    
\end{figure}

The sample is placed inside a closed-cycle cryostat operating at 1.6 K. Optical excitation is performed as illustrated in Fig.~\ref{fig:1}(a), where a resonant excitation laser beam is focused onto a waveguide by an objective positioned above the sample. The emitted photons from the QD are coupled off-chip via a focusing grating coupler and collected via the same objective.
Resonant excitation of a single QD is performed by scanning both the frequency of a continuous-wave (CW) laser and the applied bias voltage. The resulting fluorescence is plotted in Fig.~\ref{fig:2}(a).
Two bright lines are observed, both tuning with a DC Stark rate of $\Delta S = (160 \pm 30) \hspace{0.1cm} \mathrm{GHz/V}$. The two lines are separate in frequency by approximately 6 GHz, consistent with the typical fine-structure splitting between the two transition dipoles of a neutral exciton. 

A narrow optical linewidth of $(466.2 \pm 9.8)$ MHz (or $1.93 \pm 0.04$ $\mu$eV) is observed, as shown in Fig.~\ref{fig:2}(b). The measured linewidths throughout the sample (cf. Supporting Information) are comparable to those obtained from QDs grown under identical conditions in unprocessed GaAs \cite{lobl2017narrow}, i.e. in the range of 2--3.5 $\mu$eV, proving that the fabrication process did not introduce further spectral diffusion. 
Fig.~\ref{fig:2}(c) reports the time-resolved luminescence of the same QD, under pulsed resonant excitation. A single-exponential decay rate of $\gamma$ of (1.52 $\pm$ 0.01) ns$^{-1}$ is observed, corresponding to a natural linewidth $\gamma/(2\pi) = (247 \pm 38)$ MHz, which indicates that the measured linewidth is less than a factor of 2 broader than the Fourier limit. 
We note that the single-mode nanobeam waveguides employed in this work do not provide any Purcell enhancement to suppress residual dephasing as in unprocessed samples. Embedding the sources in nanocavities or slow-light photonic crystal waveguides is likely to reduce the broadening further, approaching the Fourier-transform limit \cite{pedersen2020near}. 

\begin{figure}[h!tbp]
    \centering
    \includegraphics{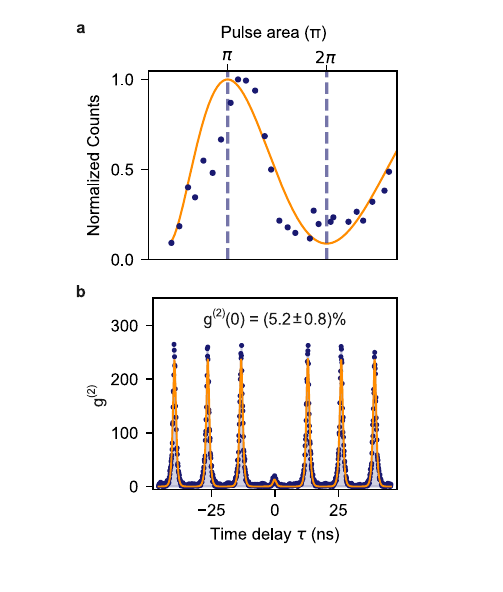}
    \caption{\textbf{Deterministic Generation of Single Photons.} 
    (a) Pulsed resonant fluorescence intensity is measured using a mode-locked laser with a repetition rate of 76 MHz and a pulse duration of approximately 20 ps, as a function of the pulse area. 
    (b) The single-photon purity measured at a $\pi$ pulse excitation, giving $g^{(2)}(0) = (5.2 \pm 0.8) \%$. The scattered points represent experimental data, while the solid lines correspond to fitted curves.}
    \label{fig:3}
\end{figure}

To verify the coherent control of the QDs and their operation as single-photon sources, we performed resonant pulsed excitation using a mode-locked Ti:Sapphire laser. The laser pulses were temporally stretched using a pulse shaper to increase the pulse duration, thereby reducing spectral broadening and minimizing laser leakage into the detection path.
The QD emission intensity as a function of excitation power exhibits clear Rabi oscillations, indicating coherent light–matter interaction. The experimental data were fitted using a damped Rabi oscillation model \cite{kosugi2005theory} and showed in Fig.~\ref{fig:3}(a).  The damping of the emission is attributed to small dephasing as well as contribution from neighboring quantum dots. As the structure is not optimized for off-chip collection with gratings, the count-rate at $\pi$-pulse is low ($\sim 1$ kHz) compared to suspended devices, which explains the relatively noisy response of Rabi oscillations. However,  designing optimized gratings and silica cladding thickness, or employing edge-coupling, is expected to increase the count rate drastically.

Finally, to confirm the single-photon nature of the quantum dot emission,  second-order correlation measurements under $\pi$-pulse excitation have been performed at bias voltage of 1.4 V and shown in Fig.~\ref{fig:3}(b). The laser background was suppressed using polarization-based filtering optics. The pronounced suppression of the zero-delay peak ($g^{(2)}(0) = (5.2 \pm 0.8)\%$) proves strong anti-bunching, with small multi-photon contribution attributed to imperfect extinction of the laser background. Additional information and details about two-photon interference and characterization of coherent operation under continuous-wave illumination is provided in the Supporting Information.

\section{Heterogeneous integration}

\begin{figure*}[h!tbp]
    \centering
    \includegraphics[width=17cm]{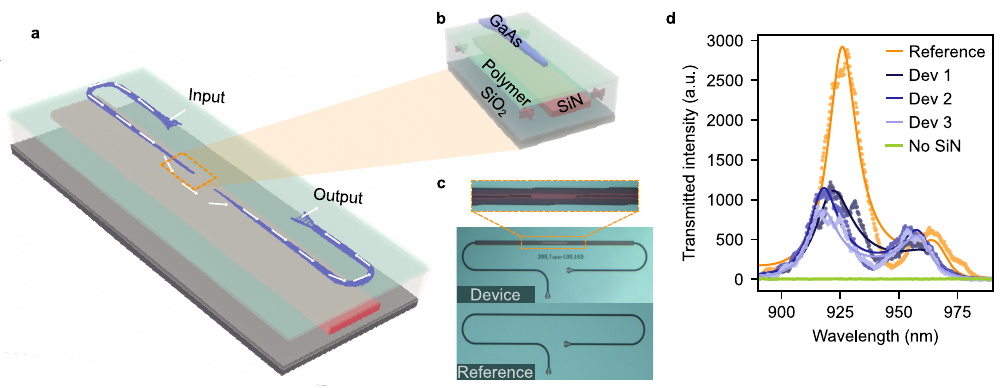}
    \caption{\textbf{Overview of Heterogeneous Integration of GaAs Waveguides with SiN.}
    (a) Schematic of the heterogeneously integrated photonic chip, where the optical mode is transferred from GaAs waveguide to the underlying SiN waveguide and back. 
    (b) Cross-sectional schematic of the device. GaAs waveguides are tapered to adiabatically transfer the optical mode to the underlying SiN waveguides. Alignment markers in the SiN layer (visible as red crosses around the waveguide) enable precise positioning of the GaAs waveguides on top.
    (c) Optical microscope images of fabricated devices. The reference waveguide is continuous (no taper structure), allowing measurement of transmission without coupling to SiN. The device shown in the image above includes a tapered end, which enables coupling from GaAs to SiN waveguides
    (d) Optical characterization of the fabricated devices. Transmission spectra are shown for the reference waveguide (no coupling to SiN) and three devices where coupling to SiN occurs. The “No SiN” case refers to a GaAs waveguide with a tapered end, but without a SiN waveguide beneath.
    }
    \label{fig:4}
\end{figure*}
To demonstrate the integration capability of the GaAs-on-insulator platform, we applied the same die-to-die bonding process described above to a SiN target chip fabricated by the commercial foundry LioniX International. The target chip features an array of straight SiN waveguides with a nominal propagation loss below 0.1 dB/cm, clad in low-pressure chemical vapor deposition (LPCVD) SiO$_2$. To ensure high-quality bonding, the SiN wafer underwent chemical mechanical planarization (CMP), resulting in a final silica thickness of $\sim200$ nm.

Fig.~\ref{fig:4}(b) shows the coupling strategy between GaAs and SiN waveguides. We use a two-step linear taper to enable adiabatic mode transfer between the two layers. Numerical simulations predict that full power transfer is possible when the GaAs taper tip width is below 110 nm, which requires electron-beam lithography and accurate alignment to predefined markers on the SiN chip. Although the visibility of these markers in silica is low during e-beam alignment, we achieve robust and reproducible positioning with an accuracy of $<$100 nm across the sample.  

We evaluated the insertion loss of the tapers by fabricating back-to-back couplers as illustrated in Fig.~\ref{fig:4}(a). Each test structure consists of two identical GaAs tapers facing each other on top of a SiN waveguide. The transmission through the two tapers is measured at cryogenic temperatures using a broadband supercontinuum source and compared with a reference device without tapers, where light remains confined to the GaAs waveguide along the path. Optical microscope images of the two devices are shown in Fig.~\ref{fig:4}(c). To reduce variations caused by the grating couplers and suppress stray-light cross-talk, the input and output gratings were cross-polarized and positioned at a fixed relative distance for all devices. Fig.~\ref{fig:4}(d) presents the transmission spectra for three nominally identical taper devices and a reference structure. To verify that light couples through the SiN waveguide rather than scattering directly from taper to taper, we fabricated a control device without the underlying SiN waveguide (labeled ``No SiN'' in the figure), which exhibited no measurable transmission. 

The broad interference fringes observed in the transmission spectra (free-spectral range of 35 nm) arises from reflections between the GaAs grating and the Si substrate, separated by a $\sim 8$ $\mu$m silica layer. To extract the insertion loss, we fitted the measured transmission curves with a Gaussian profile modulated by a Fabry-P\`{e}rot response (see Supporting Information). From the peak values of these fits, we estimate a total insertion loss of $(-3.7\pm0.3)$~dB averaged over the three devices, corresponding to a coupling efficiency of $(65\pm5)\%$  for each GaAs-to-SiN coupler. Further improvements are expected through optimized taper geometries and higher alignment accuracy, which lie beyond the scope of this work.  

\section{Conclusions}
We have demonstrated a coherent and electrically tunable single-photon source based on a GaAs quantum dot (QD) embedded in a waveguide-on-insulator platform that is fully compatible with low-loss SiN photonic integrated circuits. Embedding the QDs in a diode structure enables narrow optical linewidth and resonant excitation, preserving high coherence after integration.
The die-to-die bonding process offers a practical and relatively simple fabrication flow that can be scaled to parallel device production and, when combined with epitaxial lift-off techniques, could allow GaAs substrate reuse and significantly reduce manufacturing costs \cite{cheng2013epitaxial}. These results mark an important step towards heterogeneously integrated quantum photonic platforms that combine deterministic photon sources with mature, foundry-ready photonic circuits. Such platforms are essential for building quantum networks, implementing secure communication protocols, and enabling large-scale photonic quantum computing architectures. While resonant excitation has been successfully demonstrated here, future work should focus on improving photon extraction efficiency , optimizing coupling to photonic circuits, and extending the approach to other heterogeneous integration techniques. \\

\noindent\textbf{Ackowledgments}
We gratefully acknowledge Fabian Ruf and Martijn Heck for valuable discussions and their assistance with the design of the SiN target wafer. 
We acknowledge funding from the European Research Council (ERC) under the European Union’s Horizon 2020 research and innovation program (No. 949043, NANOMEQ), Danmarks Frie Forskningsfond (project QuPIC), the Danish National Research Foundation (Center of Excellence “Hy-Q,” grant number DNRF139),   Styrelsen for Forskning og Innovation (FI) (5072- 00016B QUANTECH), BMFTR QRN project 16KIS2200, QUANTERA BMFTR EQSOTIC project 16KIS2061, as well as DFG excellence cluster ML4Q project EXC 2004/1.

\noindent\textbf{Notes}
The authors declare no competing financial interest.

%%%%%%%%%%%%%%%%%%%%%%%%%%%%%%%%%%%%%
%%%%%%%%     Bibliography   %%%%%%%%%
%%%%%%%%%%%%%%%%%%%%%%%%%%%%%%%%%%%%%

\bibliography{main}

\begin{thebibliography}{41}%
\makeatletter
\providecommand \@ifxundefined [1]{%
 \@ifx{#1\undefined}
}%
\providecommand \@ifnum [1]{%
 \ifnum #1\expandafter \@firstoftwo
 \else \expandafter \@secondoftwo
 \fi
}%
\providecommand \@ifx [1]{%
 \ifx #1\expandafter \@firstoftwo
 \else \expandafter \@secondoftwo
 \fi
}%
\providecommand \natexlab [1]{#1}%
\providecommand \enquote  [1]{``#1''}%
\providecommand \bibnamefont  [1]{#1}%
\providecommand \bibfnamefont [1]{#1}%
\providecommand \citenamefont [1]{#1}%
\providecommand \href@noop [0]{\@secondoftwo}%
\providecommand \href [0]{\begingroup \@sanitize@url \@href}%
\providecommand \@href[1]{\@@startlink{#1}\@@href}%
\providecommand \@@href[1]{\endgroup#1\@@endlink}%
\providecommand \@sanitize@url [0]{\catcode `\\12\catcode `\$12\catcode `\&12\catcode `\#12\catcode `\^12\catcode `\_12\catcode `\%12\relax}%
\providecommand \@@startlink[1]{}%
\providecommand \@@endlink[0]{}%
\providecommand \url  [0]{\begingroup\@sanitize@url \@url }%
\providecommand \@url [1]{\endgroup\@href {#1}{\urlprefix }}%
\providecommand \urlprefix  [0]{URL }%
\providecommand \Eprint [0]{\href }%
\providecommand \doibase [0]{https://doi.org/}%
\providecommand \selectlanguage [0]{\@gobble}%
\providecommand \bibinfo  [0]{\@secondoftwo}%
\providecommand \bibfield  [0]{\@secondoftwo}%
\providecommand \translation [1]{[#1]}%
\providecommand \BibitemOpen [0]{}%
\providecommand \bibitemStop [0]{}%
\providecommand \bibitemNoStop [0]{.\EOS\space}%
\providecommand \EOS [0]{\spacefactor3000\relax}%
\providecommand \BibitemShut  [1]{\csname bibitem#1\endcsname}%
\let\auto@bib@innerbib\@empty
%</preamble>
\bibitem [{\citenamefont {Pinotsi}\ \emph {et~al.}(2011)\citenamefont {Pinotsi}, \citenamefont {Fallahi}, \citenamefont {Miguel-Sanchez},\ and\ \citenamefont {Imamoglu}}]{pinotsi2011resonant}%
  \BibitemOpen
  \bibfield  {author} {\bibinfo {author} {\bibfnamefont {D.}~\bibnamefont {Pinotsi}}, \bibinfo {author} {\bibfnamefont {P.}~\bibnamefont {Fallahi}}, \bibinfo {author} {\bibfnamefont {J.}~\bibnamefont {Miguel-Sanchez}},\ and\ \bibinfo {author} {\bibfnamefont {A.}~\bibnamefont {Imamoglu}},\ }\bibfield  {title} {\bibinfo {title} {Resonant spectroscopy on charge tunable quantum dots in photonic crystal structures},\ }\href@noop {} {\bibfield  {journal} {\bibinfo  {journal} {IEEE Journal of Quantum Electronics}\ }\textbf {\bibinfo {volume} {47}},\ \bibinfo {pages} {1371} (\bibinfo {year} {2011})}\BibitemShut {NoStop}%
\bibitem [{\citenamefont {L{\"o}bl}\ \emph {et~al.}(2017)\citenamefont {L{\"o}bl}, \citenamefont {S{\"o}llner}, \citenamefont {Javadi}, \citenamefont {Pregnolato}, \citenamefont {Schott}, \citenamefont {Midolo}, \citenamefont {Kuhlmann}, \citenamefont {Stobbe}, \citenamefont {Wieck}, \citenamefont {Lodahl} \emph {et~al.}}]{lobl2017narrow}%
  \BibitemOpen
  \bibfield  {author} {\bibinfo {author} {\bibfnamefont {M.~C.}\ \bibnamefont {L{\"o}bl}}, \bibinfo {author} {\bibfnamefont {I.}~\bibnamefont {S{\"o}llner}}, \bibinfo {author} {\bibfnamefont {A.}~\bibnamefont {Javadi}}, \bibinfo {author} {\bibfnamefont {T.}~\bibnamefont {Pregnolato}}, \bibinfo {author} {\bibfnamefont {R.}~\bibnamefont {Schott}}, \bibinfo {author} {\bibfnamefont {L.}~\bibnamefont {Midolo}}, \bibinfo {author} {\bibfnamefont {A.~V.}\ \bibnamefont {Kuhlmann}}, \bibinfo {author} {\bibfnamefont {S.}~\bibnamefont {Stobbe}}, \bibinfo {author} {\bibfnamefont {A.~D.}\ \bibnamefont {Wieck}}, \bibinfo {author} {\bibfnamefont {P.}~\bibnamefont {Lodahl}}, \emph {et~al.},\ }\bibfield  {title} {\bibinfo {title} {Narrow optical linewidths and spin pumping on charge-tunable close-to-surface self-assembled quantum dots in an ultrathin diode},\ }\href@noop {} {\bibfield  {journal} {\bibinfo  {journal} {Physical Review B}\ }\textbf {\bibinfo {volume} {96}},\ \bibinfo {pages} {165440} (\bibinfo {year}
  {2017})}\BibitemShut {NoStop}%
\bibitem [{\citenamefont {Zhai}\ \emph {et~al.}(2020)\citenamefont {Zhai}, \citenamefont {L{\"o}bl}, \citenamefont {Nguyen}, \citenamefont {Ritzmann}, \citenamefont {Javadi}, \citenamefont {Spinnler}, \citenamefont {Wieck}, \citenamefont {Ludwig},\ and\ \citenamefont {Warburton}}]{zhai2020low}%
  \BibitemOpen
  \bibfield  {author} {\bibinfo {author} {\bibfnamefont {L.}~\bibnamefont {Zhai}}, \bibinfo {author} {\bibfnamefont {M.~C.}\ \bibnamefont {L{\"o}bl}}, \bibinfo {author} {\bibfnamefont {G.~N.}\ \bibnamefont {Nguyen}}, \bibinfo {author} {\bibfnamefont {J.}~\bibnamefont {Ritzmann}}, \bibinfo {author} {\bibfnamefont {A.}~\bibnamefont {Javadi}}, \bibinfo {author} {\bibfnamefont {C.}~\bibnamefont {Spinnler}}, \bibinfo {author} {\bibfnamefont {A.~D.}\ \bibnamefont {Wieck}}, \bibinfo {author} {\bibfnamefont {A.}~\bibnamefont {Ludwig}},\ and\ \bibinfo {author} {\bibfnamefont {R.~J.}\ \bibnamefont {Warburton}},\ }\bibfield  {title} {\bibinfo {title} {Low-noise gaas quantum dots for quantum photonics},\ }\href@noop {} {\bibfield  {journal} {\bibinfo  {journal} {Nature communications}\ }\textbf {\bibinfo {volume} {11}},\ \bibinfo {pages} {4745} (\bibinfo {year} {2020})}\BibitemShut {NoStop}%
\bibitem [{\citenamefont {Kuhlmann}\ \emph {et~al.}(2015)\citenamefont {Kuhlmann}, \citenamefont {Prechtel}, \citenamefont {Houel}, \citenamefont {Ludwig}, \citenamefont {Reuter}, \citenamefont {Wieck},\ and\ \citenamefont {Warburton}}]{kuhlmann2015transform}%
  \BibitemOpen
  \bibfield  {author} {\bibinfo {author} {\bibfnamefont {A.~V.}\ \bibnamefont {Kuhlmann}}, \bibinfo {author} {\bibfnamefont {J.~H.}\ \bibnamefont {Prechtel}}, \bibinfo {author} {\bibfnamefont {J.}~\bibnamefont {Houel}}, \bibinfo {author} {\bibfnamefont {A.}~\bibnamefont {Ludwig}}, \bibinfo {author} {\bibfnamefont {D.}~\bibnamefont {Reuter}}, \bibinfo {author} {\bibfnamefont {A.~D.}\ \bibnamefont {Wieck}},\ and\ \bibinfo {author} {\bibfnamefont {R.~J.}\ \bibnamefont {Warburton}},\ }\bibfield  {title} {\bibinfo {title} {Transform-limited single photons from a single quantum dot},\ }\href@noop {} {\bibfield  {journal} {\bibinfo  {journal} {Nature communications}\ }\textbf {\bibinfo {volume} {6}},\ \bibinfo {pages} {8204} (\bibinfo {year} {2015})}\BibitemShut {NoStop}%
\bibitem [{\citenamefont {Lodahl}\ \emph {et~al.}(2015)\citenamefont {Lodahl}, \citenamefont {Mahmoodian},\ and\ \citenamefont {Stobbe}}]{lodahl2015interfacing}%
  \BibitemOpen
  \bibfield  {author} {\bibinfo {author} {\bibfnamefont {P.}~\bibnamefont {Lodahl}}, \bibinfo {author} {\bibfnamefont {S.}~\bibnamefont {Mahmoodian}},\ and\ \bibinfo {author} {\bibfnamefont {S.}~\bibnamefont {Stobbe}},\ }\bibfield  {title} {\bibinfo {title} {Interfacing single photons and single quantum dots with photonic nanostructures},\ }\href@noop {} {\bibfield  {journal} {\bibinfo  {journal} {Reviews of Modern Physics}\ }\textbf {\bibinfo {volume} {87}},\ \bibinfo {pages} {347} (\bibinfo {year} {2015})}\BibitemShut {NoStop}%
\bibitem [{\citenamefont {Arakawa}\ and\ \citenamefont {Holmes}(2020)}]{arakawa2020progress}%
  \BibitemOpen
  \bibfield  {author} {\bibinfo {author} {\bibfnamefont {Y.}~\bibnamefont {Arakawa}}\ and\ \bibinfo {author} {\bibfnamefont {M.~J.}\ \bibnamefont {Holmes}},\ }\bibfield  {title} {\bibinfo {title} {Progress in quantum-dot single photon sources for quantum information technologies: A broad spectrum overview},\ }\href@noop {} {\bibfield  {journal} {\bibinfo  {journal} {Applied Physics Reviews}\ }\textbf {\bibinfo {volume} {7}} (\bibinfo {year} {2020})}\BibitemShut {NoStop}%
\bibitem [{\citenamefont {Heindel}\ \emph {et~al.}(2023)\citenamefont {Heindel}, \citenamefont {Kim}, \citenamefont {Gregersen}, \citenamefont {Rastelli},\ and\ \citenamefont {Reitzenstein}}]{heindel2023quantum}%
  \BibitemOpen
  \bibfield  {author} {\bibinfo {author} {\bibfnamefont {T.}~\bibnamefont {Heindel}}, \bibinfo {author} {\bibfnamefont {J.-H.}\ \bibnamefont {Kim}}, \bibinfo {author} {\bibfnamefont {N.}~\bibnamefont {Gregersen}}, \bibinfo {author} {\bibfnamefont {A.}~\bibnamefont {Rastelli}},\ and\ \bibinfo {author} {\bibfnamefont {S.}~\bibnamefont {Reitzenstein}},\ }\bibfield  {title} {\bibinfo {title} {Quantum dots for photonic quantum information technology},\ }\href@noop {} {\bibfield  {journal} {\bibinfo  {journal} {Advances in Optics and Photonics}\ }\textbf {\bibinfo {volume} {15}},\ \bibinfo {pages} {613} (\bibinfo {year} {2023})}\BibitemShut {NoStop}%
\bibitem [{\citenamefont {Uppu}\ \emph {et~al.}(2020)\citenamefont {Uppu}, \citenamefont {Pedersen}, \citenamefont {Wang}, \citenamefont {Olesen}, \citenamefont {Papon}, \citenamefont {Zhou}, \citenamefont {Midolo}, \citenamefont {Scholz}, \citenamefont {Wieck}, \citenamefont {Ludwig} \emph {et~al.}}]{uppu2020scalable}%
  \BibitemOpen
  \bibfield  {author} {\bibinfo {author} {\bibfnamefont {R.}~\bibnamefont {Uppu}}, \bibinfo {author} {\bibfnamefont {F.~T.}\ \bibnamefont {Pedersen}}, \bibinfo {author} {\bibfnamefont {Y.}~\bibnamefont {Wang}}, \bibinfo {author} {\bibfnamefont {C.~T.}\ \bibnamefont {Olesen}}, \bibinfo {author} {\bibfnamefont {C.}~\bibnamefont {Papon}}, \bibinfo {author} {\bibfnamefont {X.}~\bibnamefont {Zhou}}, \bibinfo {author} {\bibfnamefont {L.}~\bibnamefont {Midolo}}, \bibinfo {author} {\bibfnamefont {S.}~\bibnamefont {Scholz}}, \bibinfo {author} {\bibfnamefont {A.~D.}\ \bibnamefont {Wieck}}, \bibinfo {author} {\bibfnamefont {A.}~\bibnamefont {Ludwig}}, \emph {et~al.},\ }\bibfield  {title} {\bibinfo {title} {Scalable integrated single-photon source},\ }\href@noop {} {\bibfield  {journal} {\bibinfo  {journal} {Science advances}\ }\textbf {\bibinfo {volume} {6}},\ \bibinfo {pages} {eabc8268} (\bibinfo {year} {2020})}\BibitemShut {NoStop}%
\bibitem [{\citenamefont {Tomm}\ \emph {et~al.}(2021)\citenamefont {Tomm}, \citenamefont {Javadi}, \citenamefont {Antoniadis}, \citenamefont {Najer}, \citenamefont {L{\"o}bl}, \citenamefont {Korsch}, \citenamefont {Schott}, \citenamefont {Valentin}, \citenamefont {Wieck}, \citenamefont {Ludwig} \emph {et~al.}}]{tomm2021bright}%
  \BibitemOpen
  \bibfield  {author} {\bibinfo {author} {\bibfnamefont {N.}~\bibnamefont {Tomm}}, \bibinfo {author} {\bibfnamefont {A.}~\bibnamefont {Javadi}}, \bibinfo {author} {\bibfnamefont {N.~O.}\ \bibnamefont {Antoniadis}}, \bibinfo {author} {\bibfnamefont {D.}~\bibnamefont {Najer}}, \bibinfo {author} {\bibfnamefont {M.~C.}\ \bibnamefont {L{\"o}bl}}, \bibinfo {author} {\bibfnamefont {A.~R.}\ \bibnamefont {Korsch}}, \bibinfo {author} {\bibfnamefont {R.}~\bibnamefont {Schott}}, \bibinfo {author} {\bibfnamefont {S.~R.}\ \bibnamefont {Valentin}}, \bibinfo {author} {\bibfnamefont {A.~D.}\ \bibnamefont {Wieck}}, \bibinfo {author} {\bibfnamefont {A.}~\bibnamefont {Ludwig}}, \emph {et~al.},\ }\bibfield  {title} {\bibinfo {title} {A bright and fast source of coherent single photons},\ }\href@noop {} {\bibfield  {journal} {\bibinfo  {journal} {Nature Nanotechnology}\ }\textbf {\bibinfo {volume} {16}},\ \bibinfo {pages} {399} (\bibinfo {year} {2021})}\BibitemShut {NoStop}%
\bibitem [{\citenamefont {Liu}\ \emph {et~al.}(2018{\natexlab{a}})\citenamefont {Liu}, \citenamefont {Brash}, \citenamefont {O’Hara}, \citenamefont {Martins}, \citenamefont {Phillips}, \citenamefont {Coles}, \citenamefont {Royall}, \citenamefont {Clarke}, \citenamefont {Bentham}, \citenamefont {Prtljaga} \emph {et~al.}}]{liu2018high}%
  \BibitemOpen
  \bibfield  {author} {\bibinfo {author} {\bibfnamefont {F.}~\bibnamefont {Liu}}, \bibinfo {author} {\bibfnamefont {A.~J.}\ \bibnamefont {Brash}}, \bibinfo {author} {\bibfnamefont {J.}~\bibnamefont {O’Hara}}, \bibinfo {author} {\bibfnamefont {L.~M.}\ \bibnamefont {Martins}}, \bibinfo {author} {\bibfnamefont {C.~L.}\ \bibnamefont {Phillips}}, \bibinfo {author} {\bibfnamefont {R.~J.}\ \bibnamefont {Coles}}, \bibinfo {author} {\bibfnamefont {B.}~\bibnamefont {Royall}}, \bibinfo {author} {\bibfnamefont {E.}~\bibnamefont {Clarke}}, \bibinfo {author} {\bibfnamefont {C.}~\bibnamefont {Bentham}}, \bibinfo {author} {\bibfnamefont {N.}~\bibnamefont {Prtljaga}}, \emph {et~al.},\ }\bibfield  {title} {\bibinfo {title} {High purcell factor generation of indistinguishable on-chip single photons},\ }\href@noop {} {\bibfield  {journal} {\bibinfo  {journal} {Nature nanotechnology}\ }\textbf {\bibinfo {volume} {13}},\ \bibinfo {pages} {835} (\bibinfo {year} {2018}{\natexlab{a}})}\BibitemShut {NoStop}%
\bibitem [{\citenamefont {Arcari}\ \emph {et~al.}(2014)\citenamefont {Arcari}, \citenamefont {S{\"o}llner}, \citenamefont {Javadi}, \citenamefont {Lindskov~Hansen}, \citenamefont {Mahmoodian}, \citenamefont {Liu}, \citenamefont {Thyrrestrup}, \citenamefont {Lee}, \citenamefont {Song}, \citenamefont {Stobbe} \emph {et~al.}}]{arcari2014near}%
  \BibitemOpen
  \bibfield  {author} {\bibinfo {author} {\bibfnamefont {M.}~\bibnamefont {Arcari}}, \bibinfo {author} {\bibfnamefont {I.}~\bibnamefont {S{\"o}llner}}, \bibinfo {author} {\bibfnamefont {A.}~\bibnamefont {Javadi}}, \bibinfo {author} {\bibfnamefont {S.}~\bibnamefont {Lindskov~Hansen}}, \bibinfo {author} {\bibfnamefont {S.}~\bibnamefont {Mahmoodian}}, \bibinfo {author} {\bibfnamefont {J.}~\bibnamefont {Liu}}, \bibinfo {author} {\bibfnamefont {H.}~\bibnamefont {Thyrrestrup}}, \bibinfo {author} {\bibfnamefont {E.~H.}\ \bibnamefont {Lee}}, \bibinfo {author} {\bibfnamefont {J.~D.}\ \bibnamefont {Song}}, \bibinfo {author} {\bibfnamefont {S.}~\bibnamefont {Stobbe}}, \emph {et~al.},\ }\bibfield  {title} {\bibinfo {title} {Near-unity coupling efficiency of a quantum emitter to a photonic crystal waveguide},\ }\href@noop {} {\bibfield  {journal} {\bibinfo  {journal} {Physical review letters}\ }\textbf {\bibinfo {volume} {113}},\ \bibinfo {pages} {093603} (\bibinfo {year} {2014})}\BibitemShut {NoStop}%
\bibitem [{\citenamefont {Kir{\v{s}}ansk{\.e}}\ \emph {et~al.}(2017)\citenamefont {Kir{\v{s}}ansk{\.e}}, \citenamefont {Thyrrestrup}, \citenamefont {Daveau}, \citenamefont {Dree{\ss}en}, \citenamefont {Pregnolato}, \citenamefont {Midolo}, \citenamefont {Tighineanu}, \citenamefont {Javadi}, \citenamefont {Stobbe}, \citenamefont {Schott} \emph {et~al.}}]{kirvsanske2017indistinguishable}%
  \BibitemOpen
  \bibfield  {author} {\bibinfo {author} {\bibfnamefont {G.}~\bibnamefont {Kir{\v{s}}ansk{\.e}}}, \bibinfo {author} {\bibfnamefont {H.}~\bibnamefont {Thyrrestrup}}, \bibinfo {author} {\bibfnamefont {R.~S.}\ \bibnamefont {Daveau}}, \bibinfo {author} {\bibfnamefont {C.~L.}\ \bibnamefont {Dree{\ss}en}}, \bibinfo {author} {\bibfnamefont {T.}~\bibnamefont {Pregnolato}}, \bibinfo {author} {\bibfnamefont {L.}~\bibnamefont {Midolo}}, \bibinfo {author} {\bibfnamefont {P.}~\bibnamefont {Tighineanu}}, \bibinfo {author} {\bibfnamefont {A.}~\bibnamefont {Javadi}}, \bibinfo {author} {\bibfnamefont {S.}~\bibnamefont {Stobbe}}, \bibinfo {author} {\bibfnamefont {R.}~\bibnamefont {Schott}}, \emph {et~al.},\ }\bibfield  {title} {\bibinfo {title} {Indistinguishable and efficient single photons from a quantum dot in a planar nanobeam waveguide},\ }\href@noop {} {\bibfield  {journal} {\bibinfo  {journal} {Physical Review B}\ }\textbf {\bibinfo {volume} {96}},\ \bibinfo {pages} {165306} (\bibinfo {year} {2017})}\BibitemShut
  {NoStop}%
\bibitem [{\citenamefont {Morais}\ \emph {et~al.}(2017)\citenamefont {Morais}, \citenamefont {Roland}, \citenamefont {Ravaro}, \citenamefont {Hease}, \citenamefont {Lema{\^\i}tre}, \citenamefont {Gomez}, \citenamefont {Wabnitz}, \citenamefont {De~Rosa}, \citenamefont {Favero},\ and\ \citenamefont {Leo}}]{morais2017directionally}%
  \BibitemOpen
  \bibfield  {author} {\bibinfo {author} {\bibfnamefont {N.}~\bibnamefont {Morais}}, \bibinfo {author} {\bibfnamefont {I.}~\bibnamefont {Roland}}, \bibinfo {author} {\bibfnamefont {M.}~\bibnamefont {Ravaro}}, \bibinfo {author} {\bibfnamefont {W.}~\bibnamefont {Hease}}, \bibinfo {author} {\bibfnamefont {A.}~\bibnamefont {Lema{\^\i}tre}}, \bibinfo {author} {\bibfnamefont {C.}~\bibnamefont {Gomez}}, \bibinfo {author} {\bibfnamefont {S.}~\bibnamefont {Wabnitz}}, \bibinfo {author} {\bibfnamefont {M.}~\bibnamefont {De~Rosa}}, \bibinfo {author} {\bibfnamefont {I.}~\bibnamefont {Favero}},\ and\ \bibinfo {author} {\bibfnamefont {G.}~\bibnamefont {Leo}},\ }\bibfield  {title} {\bibinfo {title} {Directionally induced quasi-phase matching in homogeneous algaas waveguides},\ }\href@noop {} {\bibfield  {journal} {\bibinfo  {journal} {Optics Letters}\ }\textbf {\bibinfo {volume} {42}},\ \bibinfo {pages} {4287} (\bibinfo {year} {2017})}\BibitemShut {NoStop}%
\bibitem [{\citenamefont {Zhou}\ \emph {et~al.}(2024)\citenamefont {Zhou}, \citenamefont {Yang}, \citenamefont {Wang}, \citenamefont {Koulas-Simos}, \citenamefont {Palekar}, \citenamefont {Limame}, \citenamefont {Li}, \citenamefont {Liu}, \citenamefont {Ni}, \citenamefont {Niu} \emph {et~al.}}]{zhou2024gaas}%
  \BibitemOpen
  \bibfield  {author} {\bibinfo {author} {\bibfnamefont {Y.}~\bibnamefont {Zhou}}, \bibinfo {author} {\bibfnamefont {Y.}~\bibnamefont {Yang}}, \bibinfo {author} {\bibfnamefont {Y.}~\bibnamefont {Wang}}, \bibinfo {author} {\bibfnamefont {A.}~\bibnamefont {Koulas-Simos}}, \bibinfo {author} {\bibfnamefont {C.~C.}\ \bibnamefont {Palekar}}, \bibinfo {author} {\bibfnamefont {I.}~\bibnamefont {Limame}}, \bibinfo {author} {\bibfnamefont {S.}~\bibnamefont {Li}}, \bibinfo {author} {\bibfnamefont {H.}~\bibnamefont {Liu}}, \bibinfo {author} {\bibfnamefont {H.}~\bibnamefont {Ni}}, \bibinfo {author} {\bibfnamefont {Z.}~\bibnamefont {Niu}}, \emph {et~al.},\ }\bibfield  {title} {\bibinfo {title} {Gaas-on-insulator ridge waveguide nanobeam cavities with integrated inas quantum dots},\ }\href@noop {} {\bibfield  {journal} {\bibinfo  {journal} {Materials for Quantum Technology}\ }\textbf {\bibinfo {volume} {4}},\ \bibinfo {pages} {025403} (\bibinfo {year} {2024})}\BibitemShut {NoStop}%
\bibitem [{\citenamefont {Chang}\ \emph {et~al.}(2018)\citenamefont {Chang}, \citenamefont {Boes}, \citenamefont {Guo}, \citenamefont {Spencer}, \citenamefont {Kennedy}, \citenamefont {Peters}, \citenamefont {Volet}, \citenamefont {Chiles}, \citenamefont {Kowligy}, \citenamefont {Nader} \emph {et~al.}}]{chang2018heterogeneously}%
  \BibitemOpen
  \bibfield  {author} {\bibinfo {author} {\bibfnamefont {L.}~\bibnamefont {Chang}}, \bibinfo {author} {\bibfnamefont {A.}~\bibnamefont {Boes}}, \bibinfo {author} {\bibfnamefont {X.}~\bibnamefont {Guo}}, \bibinfo {author} {\bibfnamefont {D.~T.}\ \bibnamefont {Spencer}}, \bibinfo {author} {\bibfnamefont {M.}~\bibnamefont {Kennedy}}, \bibinfo {author} {\bibfnamefont {J.~D.}\ \bibnamefont {Peters}}, \bibinfo {author} {\bibfnamefont {N.}~\bibnamefont {Volet}}, \bibinfo {author} {\bibfnamefont {J.}~\bibnamefont {Chiles}}, \bibinfo {author} {\bibfnamefont {A.}~\bibnamefont {Kowligy}}, \bibinfo {author} {\bibfnamefont {N.}~\bibnamefont {Nader}}, \emph {et~al.},\ }\bibfield  {title} {\bibinfo {title} {Heterogeneously integrated gaas waveguides on insulator for efficient frequency conversion},\ }\href@noop {} {\bibfield  {journal} {\bibinfo  {journal} {Laser \& Photonics Reviews}\ }\textbf {\bibinfo {volume} {12}},\ \bibinfo {pages} {1800149} (\bibinfo {year} {2018})}\BibitemShut {NoStop}%
\bibitem [{\citenamefont {Stanton}\ \emph {et~al.}(2020)\citenamefont {Stanton}, \citenamefont {Chiles}, \citenamefont {Nader}, \citenamefont {Moody}, \citenamefont {Volet}, \citenamefont {Chang}, \citenamefont {Bowers}, \citenamefont {Woo~Nam},\ and\ \citenamefont {Mirin}}]{stanton2020efficient}%
  \BibitemOpen
  \bibfield  {author} {\bibinfo {author} {\bibfnamefont {E.~J.}\ \bibnamefont {Stanton}}, \bibinfo {author} {\bibfnamefont {J.}~\bibnamefont {Chiles}}, \bibinfo {author} {\bibfnamefont {N.}~\bibnamefont {Nader}}, \bibinfo {author} {\bibfnamefont {G.}~\bibnamefont {Moody}}, \bibinfo {author} {\bibfnamefont {N.}~\bibnamefont {Volet}}, \bibinfo {author} {\bibfnamefont {L.}~\bibnamefont {Chang}}, \bibinfo {author} {\bibfnamefont {J.~E.}\ \bibnamefont {Bowers}}, \bibinfo {author} {\bibfnamefont {S.}~\bibnamefont {Woo~Nam}},\ and\ \bibinfo {author} {\bibfnamefont {R.~P.}\ \bibnamefont {Mirin}},\ }\bibfield  {title} {\bibinfo {title} {Efficient second harmonic generation in nanophotonic gaas-on-insulator waveguides},\ }\href@noop {} {\bibfield  {journal} {\bibinfo  {journal} {Optics express}\ }\textbf {\bibinfo {volume} {28}},\ \bibinfo {pages} {9521} (\bibinfo {year} {2020})}\BibitemShut {NoStop}%
\bibitem [{\citenamefont {Rigal}\ \emph {et~al.}(2017)\citenamefont {Rigal}, \citenamefont {Joanesarson}, \citenamefont {Lyasota}, \citenamefont {Jarlov}, \citenamefont {Dwir}, \citenamefont {Rudra}, \citenamefont {Kulkova},\ and\ \citenamefont {Kapon}}]{rigal2017propagation}%
  \BibitemOpen
  \bibfield  {author} {\bibinfo {author} {\bibfnamefont {B.}~\bibnamefont {Rigal}}, \bibinfo {author} {\bibfnamefont {K.}~\bibnamefont {Joanesarson}}, \bibinfo {author} {\bibfnamefont {A.}~\bibnamefont {Lyasota}}, \bibinfo {author} {\bibfnamefont {C.}~\bibnamefont {Jarlov}}, \bibinfo {author} {\bibfnamefont {B.}~\bibnamefont {Dwir}}, \bibinfo {author} {\bibfnamefont {A.}~\bibnamefont {Rudra}}, \bibinfo {author} {\bibfnamefont {I.}~\bibnamefont {Kulkova}},\ and\ \bibinfo {author} {\bibfnamefont {E.}~\bibnamefont {Kapon}},\ }\bibfield  {title} {\bibinfo {title} {Propagation losses in photonic crystal waveguides: effects of band tail absorption and waveguide dispersion},\ }\href@noop {} {\bibfield  {journal} {\bibinfo  {journal} {Optics Express}\ }\textbf {\bibinfo {volume} {25}},\ \bibinfo {pages} {28908} (\bibinfo {year} {2017})}\BibitemShut {NoStop}%
\bibitem [{\citenamefont {Wang}\ \emph {et~al.}(2021)\citenamefont {Wang}, \citenamefont {Uppu}, \citenamefont {Zhou}, \citenamefont {Papon}, \citenamefont {Scholz}, \citenamefont {Wieck}, \citenamefont {Ludwig}, \citenamefont {Lodahl},\ and\ \citenamefont {Midolo}}]{wang2021electroabsorption}%
  \BibitemOpen
  \bibfield  {author} {\bibinfo {author} {\bibfnamefont {Y.}~\bibnamefont {Wang}}, \bibinfo {author} {\bibfnamefont {R.}~\bibnamefont {Uppu}}, \bibinfo {author} {\bibfnamefont {X.}~\bibnamefont {Zhou}}, \bibinfo {author} {\bibfnamefont {C.}~\bibnamefont {Papon}}, \bibinfo {author} {\bibfnamefont {S.}~\bibnamefont {Scholz}}, \bibinfo {author} {\bibfnamefont {A.~D.}\ \bibnamefont {Wieck}}, \bibinfo {author} {\bibfnamefont {A.}~\bibnamefont {Ludwig}}, \bibinfo {author} {\bibfnamefont {P.}~\bibnamefont {Lodahl}},\ and\ \bibinfo {author} {\bibfnamefont {L.}~\bibnamefont {Midolo}},\ }\bibfield  {title} {\bibinfo {title} {Electroabsorption in gated gaas nanophotonic waveguides},\ }\href@noop {} {\bibfield  {journal} {\bibinfo  {journal} {Applied Physics Letters}\ }\textbf {\bibinfo {volume} {118}} (\bibinfo {year} {2021})}\BibitemShut {NoStop}%
\bibitem [{\citenamefont {Pfister}\ \emph {et~al.}(2025)\citenamefont {Pfister}, \citenamefont {Wendland}, \citenamefont {Hornung}, \citenamefont {Engel}, \citenamefont {H{\"u}ging}, \citenamefont {Herzog}, \citenamefont {Vijayan}, \citenamefont {Joos}, \citenamefont {Jung}, \citenamefont {Jetter} \emph {et~al.}}]{pfister2025telecom}%
  \BibitemOpen
  \bibfield  {author} {\bibinfo {author} {\bibfnamefont {U.}~\bibnamefont {Pfister}}, \bibinfo {author} {\bibfnamefont {D.}~\bibnamefont {Wendland}}, \bibinfo {author} {\bibfnamefont {F.}~\bibnamefont {Hornung}}, \bibinfo {author} {\bibfnamefont {L.}~\bibnamefont {Engel}}, \bibinfo {author} {\bibfnamefont {H.}~\bibnamefont {H{\"u}ging}}, \bibinfo {author} {\bibfnamefont {E.}~\bibnamefont {Herzog}}, \bibinfo {author} {\bibfnamefont {P.}~\bibnamefont {Vijayan}}, \bibinfo {author} {\bibfnamefont {R.}~\bibnamefont {Joos}}, \bibinfo {author} {\bibfnamefont {E.}~\bibnamefont {Jung}}, \bibinfo {author} {\bibfnamefont {M.}~\bibnamefont {Jetter}}, \emph {et~al.},\ }\bibfield  {title} {\bibinfo {title} {Telecom wavelength quantum dots interfaced with silicon-nitride circuits via photonic wire bonding},\ }\href@noop {} {\bibfield  {journal} {\bibinfo  {journal} {npj Nanophotonics}\ }\textbf {\bibinfo {volume} {2}},\ \bibinfo {pages} {11} (\bibinfo {year} {2025})}\BibitemShut {NoStop}%
\bibitem [{\citenamefont {Chanana}\ \emph {et~al.}(2022)\citenamefont {Chanana}, \citenamefont {Larocque}, \citenamefont {Moreira}, \citenamefont {Carolan}, \citenamefont {Guha}, \citenamefont {Melo}, \citenamefont {Anant}, \citenamefont {Song}, \citenamefont {Englund}, \citenamefont {Blumenthal} \emph {et~al.}}]{chanana2022ultra}%
  \BibitemOpen
  \bibfield  {author} {\bibinfo {author} {\bibfnamefont {A.}~\bibnamefont {Chanana}}, \bibinfo {author} {\bibfnamefont {H.}~\bibnamefont {Larocque}}, \bibinfo {author} {\bibfnamefont {R.}~\bibnamefont {Moreira}}, \bibinfo {author} {\bibfnamefont {J.}~\bibnamefont {Carolan}}, \bibinfo {author} {\bibfnamefont {B.}~\bibnamefont {Guha}}, \bibinfo {author} {\bibfnamefont {E.~G.}\ \bibnamefont {Melo}}, \bibinfo {author} {\bibfnamefont {V.}~\bibnamefont {Anant}}, \bibinfo {author} {\bibfnamefont {J.}~\bibnamefont {Song}}, \bibinfo {author} {\bibfnamefont {D.}~\bibnamefont {Englund}}, \bibinfo {author} {\bibfnamefont {D.~J.}\ \bibnamefont {Blumenthal}}, \emph {et~al.},\ }\bibfield  {title} {\bibinfo {title} {Ultra-low loss quantum photonic circuits integrated with single quantum emitters},\ }\href@noop {} {\bibfield  {journal} {\bibinfo  {journal} {Nature Communications}\ }\textbf {\bibinfo {volume} {13}},\ \bibinfo {pages} {7693} (\bibinfo {year} {2022})}\BibitemShut {NoStop}%
\bibitem [{\citenamefont {Descamps}\ \emph {et~al.}(2024)\citenamefont {Descamps}, \citenamefont {Schetelat}, \citenamefont {Gao}, \citenamefont {Poole}, \citenamefont {Dalacu}, \citenamefont {Elshaari},\ and\ \citenamefont {Zwiller}}]{descamps2024acoustic}%
  \BibitemOpen
  \bibfield  {author} {\bibinfo {author} {\bibfnamefont {T.}~\bibnamefont {Descamps}}, \bibinfo {author} {\bibfnamefont {T.}~\bibnamefont {Schetelat}}, \bibinfo {author} {\bibfnamefont {J.}~\bibnamefont {Gao}}, \bibinfo {author} {\bibfnamefont {P.~J.}\ \bibnamefont {Poole}}, \bibinfo {author} {\bibfnamefont {D.}~\bibnamefont {Dalacu}}, \bibinfo {author} {\bibfnamefont {A.~W.}\ \bibnamefont {Elshaari}},\ and\ \bibinfo {author} {\bibfnamefont {V.}~\bibnamefont {Zwiller}},\ }\bibfield  {title} {\bibinfo {title} {Acoustic modulation of individual nanowire quantum dots integrated into a hybrid thin-film lithium niobate photonic platform},\ }\href@noop {} {\bibfield  {journal} {\bibinfo  {journal} {Nano Letters}\ }\textbf {\bibinfo {volume} {24}},\ \bibinfo {pages} {12493} (\bibinfo {year} {2024})}\BibitemShut {NoStop}%
\bibitem [{\citenamefont {Katsumi}\ \emph {et~al.}(2022)\citenamefont {Katsumi}, \citenamefont {Ota}, \citenamefont {Tajiri}, \citenamefont {Iwamoto}, \citenamefont {Ranbir}, \citenamefont {Reithmaier}, \citenamefont {Benyoucef},\ and\ \citenamefont {Arakawa}}]{katsumi2022cmos}%
  \BibitemOpen
  \bibfield  {author} {\bibinfo {author} {\bibfnamefont {R.}~\bibnamefont {Katsumi}}, \bibinfo {author} {\bibfnamefont {Y.}~\bibnamefont {Ota}}, \bibinfo {author} {\bibfnamefont {T.}~\bibnamefont {Tajiri}}, \bibinfo {author} {\bibfnamefont {S.}~\bibnamefont {Iwamoto}}, \bibinfo {author} {\bibfnamefont {K.}~\bibnamefont {Ranbir}}, \bibinfo {author} {\bibfnamefont {J.~P.}\ \bibnamefont {Reithmaier}}, \bibinfo {author} {\bibfnamefont {M.}~\bibnamefont {Benyoucef}},\ and\ \bibinfo {author} {\bibfnamefont {Y.}~\bibnamefont {Arakawa}},\ }\bibfield  {title} {\bibinfo {title} {Cmos-compatible integration of telecom band inas/inp quantum-dot single-photon sources on a si chip using transfer printing},\ }\href@noop {} {\bibfield  {journal} {\bibinfo  {journal} {Applied Physics Express}\ }\textbf {\bibinfo {volume} {16}},\ \bibinfo {pages} {012004} (\bibinfo {year} {2022})}\BibitemShut {NoStop}%
\bibitem [{\citenamefont {Liang}\ \emph {et~al.}(2010)\citenamefont {Liang}, \citenamefont {Roelkens}, \citenamefont {Baets},\ and\ \citenamefont {Bowers}}]{liang2010hybrid}%
  \BibitemOpen
  \bibfield  {author} {\bibinfo {author} {\bibfnamefont {D.}~\bibnamefont {Liang}}, \bibinfo {author} {\bibfnamefont {G.}~\bibnamefont {Roelkens}}, \bibinfo {author} {\bibfnamefont {R.}~\bibnamefont {Baets}},\ and\ \bibinfo {author} {\bibfnamefont {J.~E.}\ \bibnamefont {Bowers}},\ }\bibfield  {title} {\bibinfo {title} {Hybrid integrated platforms for silicon photonics},\ }\href@noop {} {\bibfield  {journal} {\bibinfo  {journal} {Materials}\ }\textbf {\bibinfo {volume} {3}},\ \bibinfo {pages} {1782} (\bibinfo {year} {2010})}\BibitemShut {NoStop}%
\bibitem [{\citenamefont {Osada}\ \emph {et~al.}(2019)\citenamefont {Osada}, \citenamefont {Ota}, \citenamefont {Katsumi}, \citenamefont {Kakuda}, \citenamefont {Iwamoto},\ and\ \citenamefont {Arakawa}}]{osada2019strongly}%
  \BibitemOpen
  \bibfield  {author} {\bibinfo {author} {\bibfnamefont {A.}~\bibnamefont {Osada}}, \bibinfo {author} {\bibfnamefont {Y.}~\bibnamefont {Ota}}, \bibinfo {author} {\bibfnamefont {R.}~\bibnamefont {Katsumi}}, \bibinfo {author} {\bibfnamefont {M.}~\bibnamefont {Kakuda}}, \bibinfo {author} {\bibfnamefont {S.}~\bibnamefont {Iwamoto}},\ and\ \bibinfo {author} {\bibfnamefont {Y.}~\bibnamefont {Arakawa}},\ }\bibfield  {title} {\bibinfo {title} {Strongly coupled single-quantum-dot--cavity system integrated on a cmos-processed silicon photonic chip},\ }\href@noop {} {\bibfield  {journal} {\bibinfo  {journal} {Physical Review Applied}\ }\textbf {\bibinfo {volume} {11}},\ \bibinfo {pages} {024071} (\bibinfo {year} {2019})}\BibitemShut {NoStop}%
\bibitem [{\citenamefont {Kim}\ \emph {et~al.}(2017)\citenamefont {Kim}, \citenamefont {Aghaeimeibodi}, \citenamefont {Richardson}, \citenamefont {Leavitt}, \citenamefont {Englund},\ and\ \citenamefont {Waks}}]{kim2017hybrid}%
  \BibitemOpen
  \bibfield  {author} {\bibinfo {author} {\bibfnamefont {J.-H.}\ \bibnamefont {Kim}}, \bibinfo {author} {\bibfnamefont {S.}~\bibnamefont {Aghaeimeibodi}}, \bibinfo {author} {\bibfnamefont {C.~J.}\ \bibnamefont {Richardson}}, \bibinfo {author} {\bibfnamefont {R.~P.}\ \bibnamefont {Leavitt}}, \bibinfo {author} {\bibfnamefont {D.}~\bibnamefont {Englund}},\ and\ \bibinfo {author} {\bibfnamefont {E.}~\bibnamefont {Waks}},\ }\bibfield  {title} {\bibinfo {title} {Hybrid integration of solid-state quantum emitters on a silicon photonic chip},\ }\href@noop {} {\bibfield  {journal} {\bibinfo  {journal} {Nano letters}\ }\textbf {\bibinfo {volume} {17}},\ \bibinfo {pages} {7394} (\bibinfo {year} {2017})}\BibitemShut {NoStop}%
\bibitem [{\citenamefont {Chandrasekar}(2022)}]{chandrasekar2022mechanophotonics}%
  \BibitemOpen
  \bibfield  {author} {\bibinfo {author} {\bibfnamefont {R.}~\bibnamefont {Chandrasekar}},\ }\bibfield  {title} {\bibinfo {title} {Mechanophotonics--a guide to integrating microcrystals toward monolithic and hybrid all-organic photonic circuits},\ }\href@noop {} {\bibfield  {journal} {\bibinfo  {journal} {Chemical Communications}\ }\textbf {\bibinfo {volume} {58}},\ \bibinfo {pages} {3415} (\bibinfo {year} {2022})}\BibitemShut {NoStop}%
\bibitem [{\citenamefont {Larocque}\ \emph {et~al.}(2024)\citenamefont {Larocque}, \citenamefont {Buyukkaya}, \citenamefont {Errando-Herranz}, \citenamefont {Papon}, \citenamefont {Harper}, \citenamefont {Tao}, \citenamefont {Carolan}, \citenamefont {Lee}, \citenamefont {Richardson}, \citenamefont {Leake} \emph {et~al.}}]{larocque2024tunable}%
  \BibitemOpen
  \bibfield  {author} {\bibinfo {author} {\bibfnamefont {H.}~\bibnamefont {Larocque}}, \bibinfo {author} {\bibfnamefont {M.~A.}\ \bibnamefont {Buyukkaya}}, \bibinfo {author} {\bibfnamefont {C.}~\bibnamefont {Errando-Herranz}}, \bibinfo {author} {\bibfnamefont {C.}~\bibnamefont {Papon}}, \bibinfo {author} {\bibfnamefont {S.}~\bibnamefont {Harper}}, \bibinfo {author} {\bibfnamefont {M.}~\bibnamefont {Tao}}, \bibinfo {author} {\bibfnamefont {J.}~\bibnamefont {Carolan}}, \bibinfo {author} {\bibfnamefont {C.-M.}\ \bibnamefont {Lee}}, \bibinfo {author} {\bibfnamefont {C.~J.}\ \bibnamefont {Richardson}}, \bibinfo {author} {\bibfnamefont {G.~L.}\ \bibnamefont {Leake}}, \emph {et~al.},\ }\bibfield  {title} {\bibinfo {title} {Tunable quantum emitters on large-scale foundry silicon photonics},\ }\href@noop {} {\bibfield  {journal} {\bibinfo  {journal} {Nature Communications}\ }\textbf {\bibinfo {volume} {15}},\ \bibinfo {pages} {5781} (\bibinfo {year} {2024})}\BibitemShut {NoStop}%
\bibitem [{\citenamefont {Shadmani}\ \emph {et~al.}(2022)\citenamefont {Shadmani}, \citenamefont {Thomas}, \citenamefont {Liu}, \citenamefont {Papon}, \citenamefont {Heck}, \citenamefont {Volet}, \citenamefont {Scholz}, \citenamefont {Wieck}, \citenamefont {Ludwig}, \citenamefont {Lodahl} \emph {et~al.}}]{shadmani2022integration}%
  \BibitemOpen
  \bibfield  {author} {\bibinfo {author} {\bibfnamefont {A.}~\bibnamefont {Shadmani}}, \bibinfo {author} {\bibfnamefont {R.~A.}\ \bibnamefont {Thomas}}, \bibinfo {author} {\bibfnamefont {Z.}~\bibnamefont {Liu}}, \bibinfo {author} {\bibfnamefont {C.}~\bibnamefont {Papon}}, \bibinfo {author} {\bibfnamefont {M.~J.}\ \bibnamefont {Heck}}, \bibinfo {author} {\bibfnamefont {N.}~\bibnamefont {Volet}}, \bibinfo {author} {\bibfnamefont {S.}~\bibnamefont {Scholz}}, \bibinfo {author} {\bibfnamefont {A.~D.}\ \bibnamefont {Wieck}}, \bibinfo {author} {\bibfnamefont {A.}~\bibnamefont {Ludwig}}, \bibinfo {author} {\bibfnamefont {P.}~\bibnamefont {Lodahl}}, \emph {et~al.},\ }\bibfield  {title} {\bibinfo {title} {Integration of gaas waveguides on a silicon substrate for quantum photonic circuits},\ }\href@noop {} {\bibfield  {journal} {\bibinfo  {journal} {Optics Express}\ }\textbf {\bibinfo {volume} {30}},\ \bibinfo {pages} {37595} (\bibinfo {year} {2022})}\BibitemShut {NoStop}%
\bibitem [{\citenamefont {Davanco}\ \emph {et~al.}(2017)\citenamefont {Davanco}, \citenamefont {Liu}, \citenamefont {Sapienza}, \citenamefont {Zhang}, \citenamefont {De~Miranda~Cardoso}, \citenamefont {Verma}, \citenamefont {Mirin}, \citenamefont {Nam}, \citenamefont {Liu},\ and\ \citenamefont {Srinivasan}}]{davanco2017heterogeneous}%
  \BibitemOpen
  \bibfield  {author} {\bibinfo {author} {\bibfnamefont {M.}~\bibnamefont {Davanco}}, \bibinfo {author} {\bibfnamefont {J.}~\bibnamefont {Liu}}, \bibinfo {author} {\bibfnamefont {L.}~\bibnamefont {Sapienza}}, \bibinfo {author} {\bibfnamefont {C.-Z.}\ \bibnamefont {Zhang}}, \bibinfo {author} {\bibfnamefont {J.~V.}\ \bibnamefont {De~Miranda~Cardoso}}, \bibinfo {author} {\bibfnamefont {V.}~\bibnamefont {Verma}}, \bibinfo {author} {\bibfnamefont {R.}~\bibnamefont {Mirin}}, \bibinfo {author} {\bibfnamefont {S.~W.}\ \bibnamefont {Nam}}, \bibinfo {author} {\bibfnamefont {L.}~\bibnamefont {Liu}},\ and\ \bibinfo {author} {\bibfnamefont {K.}~\bibnamefont {Srinivasan}},\ }\bibfield  {title} {\bibinfo {title} {Heterogeneous integration for on-chip quantum photonic circuits with single quantum dot devices},\ }\href@noop {} {\bibfield  {journal} {\bibinfo  {journal} {Nature communications}\ }\textbf {\bibinfo {volume} {8}},\ \bibinfo {pages} {889} (\bibinfo {year} {2017})}\BibitemShut {NoStop}%
\bibitem [{\citenamefont {Schnauber}\ \emph {et~al.}(2019)\citenamefont {Schnauber}, \citenamefont {Singh}, \citenamefont {Schall}, \citenamefont {Park}, \citenamefont {Song}, \citenamefont {Rodt}, \citenamefont {Srinivasan}, \citenamefont {Reitzenstein},\ and\ \citenamefont {Davanco}}]{schnauber2019indistinguishable}%
  \BibitemOpen
  \bibfield  {author} {\bibinfo {author} {\bibfnamefont {P.}~\bibnamefont {Schnauber}}, \bibinfo {author} {\bibfnamefont {A.}~\bibnamefont {Singh}}, \bibinfo {author} {\bibfnamefont {J.}~\bibnamefont {Schall}}, \bibinfo {author} {\bibfnamefont {S.~I.}\ \bibnamefont {Park}}, \bibinfo {author} {\bibfnamefont {J.~D.}\ \bibnamefont {Song}}, \bibinfo {author} {\bibfnamefont {S.}~\bibnamefont {Rodt}}, \bibinfo {author} {\bibfnamefont {K.}~\bibnamefont {Srinivasan}}, \bibinfo {author} {\bibfnamefont {S.}~\bibnamefont {Reitzenstein}},\ and\ \bibinfo {author} {\bibfnamefont {M.}~\bibnamefont {Davanco}},\ }\bibfield  {title} {\bibinfo {title} {Indistinguishable photons from deterministically integrated single quantum dots in heterogeneous gaas/si3n4 quantum photonic circuits},\ }\href@noop {} {\bibfield  {journal} {\bibinfo  {journal} {Nano letters}\ }\textbf {\bibinfo {volume} {19}},\ \bibinfo {pages} {7164} (\bibinfo {year} {2019})}\BibitemShut {NoStop}%
\bibitem [{\citenamefont {Le~Jeannic}\ \emph {et~al.}(2022)\citenamefont {Le~Jeannic}, \citenamefont {Tiranov}, \citenamefont {Carolan}, \citenamefont {Ramos}, \citenamefont {Wang}, \citenamefont {Appel}, \citenamefont {Scholz}, \citenamefont {Wieck}, \citenamefont {Ludwig}, \citenamefont {Rotenberg} \emph {et~al.}}]{le2022dynamical}%
  \BibitemOpen
  \bibfield  {author} {\bibinfo {author} {\bibfnamefont {H.}~\bibnamefont {Le~Jeannic}}, \bibinfo {author} {\bibfnamefont {A.}~\bibnamefont {Tiranov}}, \bibinfo {author} {\bibfnamefont {J.}~\bibnamefont {Carolan}}, \bibinfo {author} {\bibfnamefont {T.}~\bibnamefont {Ramos}}, \bibinfo {author} {\bibfnamefont {Y.}~\bibnamefont {Wang}}, \bibinfo {author} {\bibfnamefont {M.~H.}\ \bibnamefont {Appel}}, \bibinfo {author} {\bibfnamefont {S.}~\bibnamefont {Scholz}}, \bibinfo {author} {\bibfnamefont {A.~D.}\ \bibnamefont {Wieck}}, \bibinfo {author} {\bibfnamefont {A.}~\bibnamefont {Ludwig}}, \bibinfo {author} {\bibfnamefont {N.}~\bibnamefont {Rotenberg}}, \emph {et~al.},\ }\bibfield  {title} {\bibinfo {title} {Dynamical photon--photon interaction mediated by a quantum emitter},\ }\href@noop {} {\bibfield  {journal} {\bibinfo  {journal} {Nature Physics}\ }\textbf {\bibinfo {volume} {18}},\ \bibinfo {pages} {1191} (\bibinfo {year} {2022})}\BibitemShut {NoStop}%
\bibitem [{\citenamefont {Meng}\ \emph {et~al.}(2024)\citenamefont {Meng}, \citenamefont {Chan}, \citenamefont {Nielsen}, \citenamefont {Appel}, \citenamefont {Liu}, \citenamefont {Wang}, \citenamefont {Bart}, \citenamefont {Wieck}, \citenamefont {Ludwig}, \citenamefont {Midolo} \emph {et~al.}}]{meng2024deterministic}%
  \BibitemOpen
  \bibfield  {author} {\bibinfo {author} {\bibfnamefont {Y.}~\bibnamefont {Meng}}, \bibinfo {author} {\bibfnamefont {M.~L.}\ \bibnamefont {Chan}}, \bibinfo {author} {\bibfnamefont {R.~B.}\ \bibnamefont {Nielsen}}, \bibinfo {author} {\bibfnamefont {M.~H.}\ \bibnamefont {Appel}}, \bibinfo {author} {\bibfnamefont {Z.}~\bibnamefont {Liu}}, \bibinfo {author} {\bibfnamefont {Y.}~\bibnamefont {Wang}}, \bibinfo {author} {\bibfnamefont {N.}~\bibnamefont {Bart}}, \bibinfo {author} {\bibfnamefont {A.~D.}\ \bibnamefont {Wieck}}, \bibinfo {author} {\bibfnamefont {A.}~\bibnamefont {Ludwig}}, \bibinfo {author} {\bibfnamefont {L.}~\bibnamefont {Midolo}}, \emph {et~al.},\ }\bibfield  {title} {\bibinfo {title} {Deterministic photon source of genuine three-qubit entanglement},\ }\href@noop {} {\bibfield  {journal} {\bibinfo  {journal} {Nature communications}\ }\textbf {\bibinfo {volume} {15}},\ \bibinfo {pages} {7774} (\bibinfo {year} {2024})}\BibitemShut {NoStop}%
\bibitem [{\citenamefont {Grim}\ \emph {et~al.}(2019)\citenamefont {Grim}, \citenamefont {Bracker}, \citenamefont {Zalalutdinov}, \citenamefont {Carter}, \citenamefont {Kozen}, \citenamefont {Kim}, \citenamefont {Kim}, \citenamefont {Mlack}, \citenamefont {Yakes}, \citenamefont {Lee} \emph {et~al.}}]{grim2019scalable}%
  \BibitemOpen
  \bibfield  {author} {\bibinfo {author} {\bibfnamefont {J.~Q.}\ \bibnamefont {Grim}}, \bibinfo {author} {\bibfnamefont {A.~S.}\ \bibnamefont {Bracker}}, \bibinfo {author} {\bibfnamefont {M.}~\bibnamefont {Zalalutdinov}}, \bibinfo {author} {\bibfnamefont {S.~G.}\ \bibnamefont {Carter}}, \bibinfo {author} {\bibfnamefont {A.~C.}\ \bibnamefont {Kozen}}, \bibinfo {author} {\bibfnamefont {M.}~\bibnamefont {Kim}}, \bibinfo {author} {\bibfnamefont {C.~S.}\ \bibnamefont {Kim}}, \bibinfo {author} {\bibfnamefont {J.~T.}\ \bibnamefont {Mlack}}, \bibinfo {author} {\bibfnamefont {M.}~\bibnamefont {Yakes}}, \bibinfo {author} {\bibfnamefont {B.}~\bibnamefont {Lee}}, \emph {et~al.},\ }\bibfield  {title} {\bibinfo {title} {Scalable in operando strain tuning in nanophotonic waveguides enabling three-quantum-dot superradiance},\ }\href@noop {} {\bibfield  {journal} {\bibinfo  {journal} {Nature materials}\ }\textbf {\bibinfo {volume} {18}},\ \bibinfo {pages} {963} (\bibinfo {year} {2019})}\BibitemShut {NoStop}%
\bibitem [{\citenamefont {Patel}\ \emph {et~al.}(2010)\citenamefont {Patel}, \citenamefont {Bennett}, \citenamefont {Farrer}, \citenamefont {Nicoll}, \citenamefont {Ritchie},\ and\ \citenamefont {Shields}}]{patel2010two}%
  \BibitemOpen
  \bibfield  {author} {\bibinfo {author} {\bibfnamefont {R.~B.}\ \bibnamefont {Patel}}, \bibinfo {author} {\bibfnamefont {A.~J.}\ \bibnamefont {Bennett}}, \bibinfo {author} {\bibfnamefont {I.}~\bibnamefont {Farrer}}, \bibinfo {author} {\bibfnamefont {C.~A.}\ \bibnamefont {Nicoll}}, \bibinfo {author} {\bibfnamefont {D.~A.}\ \bibnamefont {Ritchie}},\ and\ \bibinfo {author} {\bibfnamefont {A.~J.}\ \bibnamefont {Shields}},\ }\bibfield  {title} {\bibinfo {title} {Two-photon interference of the emission from electrically tunable remote quantum dots},\ }\href@noop {} {\bibfield  {journal} {\bibinfo  {journal} {Nature photonics}\ }\textbf {\bibinfo {volume} {4}},\ \bibinfo {pages} {632} (\bibinfo {year} {2010})}\BibitemShut {NoStop}%
\bibitem [{\citenamefont {Pedersen}\ \emph {et~al.}(2020)\citenamefont {Pedersen}, \citenamefont {Wang}, \citenamefont {Olesen}, \citenamefont {Scholz}, \citenamefont {Wieck}, \citenamefont {Ludwig}, \citenamefont {LObl}, \citenamefont {Warburton}, \citenamefont {Midolo}, \citenamefont {Uppu} \emph {et~al.}}]{pedersen2020near}%
  \BibitemOpen
  \bibfield  {author} {\bibinfo {author} {\bibfnamefont {F.~T.}\ \bibnamefont {Pedersen}}, \bibinfo {author} {\bibfnamefont {Y.}~\bibnamefont {Wang}}, \bibinfo {author} {\bibfnamefont {C.~T.}\ \bibnamefont {Olesen}}, \bibinfo {author} {\bibfnamefont {S.}~\bibnamefont {Scholz}}, \bibinfo {author} {\bibfnamefont {A.~D.}\ \bibnamefont {Wieck}}, \bibinfo {author} {\bibfnamefont {A.}~\bibnamefont {Ludwig}}, \bibinfo {author} {\bibfnamefont {M.~C.}\ \bibnamefont {LObl}}, \bibinfo {author} {\bibfnamefont {R.~J.}\ \bibnamefont {Warburton}}, \bibinfo {author} {\bibfnamefont {L.}~\bibnamefont {Midolo}}, \bibinfo {author} {\bibfnamefont {R.}~\bibnamefont {Uppu}}, \emph {et~al.},\ }\bibfield  {title} {\bibinfo {title} {Near transform-limited quantum dot linewidths in a broadband photonic crystal waveguide},\ }\href@noop {} {\bibfield  {journal} {\bibinfo  {journal} {ACS Photonics}\ }\textbf {\bibinfo {volume} {7}},\ \bibinfo {pages} {2343} (\bibinfo {year} {2020})}\BibitemShut {NoStop}%
\bibitem [{\citenamefont {Kosugi}\ \emph {et~al.}(2005)\citenamefont {Kosugi}, \citenamefont {Matsuo}, \citenamefont {Konno},\ and\ \citenamefont {Hatakenaka}}]{kosugi2005theory}%
  \BibitemOpen
  \bibfield  {author} {\bibinfo {author} {\bibfnamefont {N.}~\bibnamefont {Kosugi}}, \bibinfo {author} {\bibfnamefont {S.}~\bibnamefont {Matsuo}}, \bibinfo {author} {\bibfnamefont {K.}~\bibnamefont {Konno}},\ and\ \bibinfo {author} {\bibfnamefont {N.}~\bibnamefont {Hatakenaka}},\ }\bibfield  {title} {\bibinfo {title} {Theory of damped rabi oscillations},\ }\href@noop {} {\bibfield  {journal} {\bibinfo  {journal} {Physical Review B—Condensed Matter and Materials Physics}\ }\textbf {\bibinfo {volume} {72}},\ \bibinfo {pages} {172509} (\bibinfo {year} {2005})}\BibitemShut {NoStop}%
\bibitem [{\citenamefont {Cheng}\ \emph {et~al.}(2013)\citenamefont {Cheng}, \citenamefont {Shiu}, \citenamefont {Li}, \citenamefont {Han}, \citenamefont {Shi},\ and\ \citenamefont {Sadana}}]{cheng2013epitaxial}%
  \BibitemOpen
  \bibfield  {author} {\bibinfo {author} {\bibfnamefont {C.-W.}\ \bibnamefont {Cheng}}, \bibinfo {author} {\bibfnamefont {K.-T.}\ \bibnamefont {Shiu}}, \bibinfo {author} {\bibfnamefont {N.}~\bibnamefont {Li}}, \bibinfo {author} {\bibfnamefont {S.-J.}\ \bibnamefont {Han}}, \bibinfo {author} {\bibfnamefont {L.}~\bibnamefont {Shi}},\ and\ \bibinfo {author} {\bibfnamefont {D.~K.}\ \bibnamefont {Sadana}},\ }\bibfield  {title} {\bibinfo {title} {Epitaxial lift-off process for gallium arsenide substrate reuse and flexible electronics},\ }\href@noop {} {\bibfield  {journal} {\bibinfo  {journal} {Nature communications}\ }\textbf {\bibinfo {volume} {4}},\ \bibinfo {pages} {1577} (\bibinfo {year} {2013})}\BibitemShut {NoStop}%
\bibitem [{\citenamefont {Davan{\c{c}}o}\ \emph {et~al.}(2014)\citenamefont {Davan{\c{c}}o}, \citenamefont {Hellberg}, \citenamefont {Ates}, \citenamefont {Badolato},\ and\ \citenamefont {Srinivasan}}]{davancco2014multiple}%
  \BibitemOpen
  \bibfield  {author} {\bibinfo {author} {\bibfnamefont {M.}~\bibnamefont {Davan{\c{c}}o}}, \bibinfo {author} {\bibfnamefont {C.~S.}\ \bibnamefont {Hellberg}}, \bibinfo {author} {\bibfnamefont {S.}~\bibnamefont {Ates}}, \bibinfo {author} {\bibfnamefont {A.}~\bibnamefont {Badolato}},\ and\ \bibinfo {author} {\bibfnamefont {K.}~\bibnamefont {Srinivasan}},\ }\bibfield  {title} {\bibinfo {title} {Multiple time scale blinking in inas quantum dot single-photon sources},\ }\href@noop {} {\bibfield  {journal} {\bibinfo  {journal} {Physical Review B}\ }\textbf {\bibinfo {volume} {89}},\ \bibinfo {pages} {161303} (\bibinfo {year} {2014})}\BibitemShut {NoStop}%
\bibitem [{\citenamefont {Liu}\ \emph {et~al.}(2018{\natexlab{b}})\citenamefont {Liu}, \citenamefont {Konthasinghe}, \citenamefont {Davan{\c{c}}o}, \citenamefont {Lawall}, \citenamefont {Anant}, \citenamefont {Verma}, \citenamefont {Mirin}, \citenamefont {Nam}, \citenamefont {Song}, \citenamefont {Ma} \emph {et~al.}}]{liu2018single}%
  \BibitemOpen
  \bibfield  {author} {\bibinfo {author} {\bibfnamefont {J.}~\bibnamefont {Liu}}, \bibinfo {author} {\bibfnamefont {K.}~\bibnamefont {Konthasinghe}}, \bibinfo {author} {\bibfnamefont {M.}~\bibnamefont {Davan{\c{c}}o}}, \bibinfo {author} {\bibfnamefont {J.}~\bibnamefont {Lawall}}, \bibinfo {author} {\bibfnamefont {V.}~\bibnamefont {Anant}}, \bibinfo {author} {\bibfnamefont {V.}~\bibnamefont {Verma}}, \bibinfo {author} {\bibfnamefont {R.}~\bibnamefont {Mirin}}, \bibinfo {author} {\bibfnamefont {S.~W.}\ \bibnamefont {Nam}}, \bibinfo {author} {\bibfnamefont {J.~D.}\ \bibnamefont {Song}}, \bibinfo {author} {\bibfnamefont {B.}~\bibnamefont {Ma}}, \emph {et~al.},\ }\bibfield  {title} {\bibinfo {title} {Single self-assembled inas/gaas quantum dots in photonic nanostructures: the role of nanofabrication},\ }\href@noop {} {\bibfield  {journal} {\bibinfo  {journal} {Physical review applied}\ }\textbf {\bibinfo {volume} {9}},\ \bibinfo {pages} {064019} (\bibinfo {year} {2018}{\natexlab{b}})}\BibitemShut {NoStop}%
\bibitem [{\citenamefont {Schofield}\ \emph {et~al.}(2022)\citenamefont {Schofield}, \citenamefont {Clear}, \citenamefont {Hoggarth}, \citenamefont {Major}, \citenamefont {McCutcheon},\ and\ \citenamefont {Clark}}]{schofield2022photon}%
  \BibitemOpen
  \bibfield  {author} {\bibinfo {author} {\bibfnamefont {R.~C.}\ \bibnamefont {Schofield}}, \bibinfo {author} {\bibfnamefont {C.}~\bibnamefont {Clear}}, \bibinfo {author} {\bibfnamefont {R.~A.}\ \bibnamefont {Hoggarth}}, \bibinfo {author} {\bibfnamefont {K.~D.}\ \bibnamefont {Major}}, \bibinfo {author} {\bibfnamefont {D.~P.}\ \bibnamefont {McCutcheon}},\ and\ \bibinfo {author} {\bibfnamefont {A.~S.}\ \bibnamefont {Clark}},\ }\bibfield  {title} {\bibinfo {title} {Photon indistinguishability measurements under pulsed and continuous excitation},\ }\href@noop {} {\bibfield  {journal} {\bibinfo  {journal} {Physical Review Research}\ }\textbf {\bibinfo {volume} {4}},\ \bibinfo {pages} {013037} (\bibinfo {year} {2022})}\BibitemShut {NoStop}%
\bibitem [{\citenamefont {Loredo}\ \emph {et~al.}(2019)\citenamefont {Loredo}, \citenamefont {Ant{\'o}n}, \citenamefont {Reznychenko}, \citenamefont {Hilaire}, \citenamefont {Harouri}, \citenamefont {Millet}, \citenamefont {Ollivier}, \citenamefont {Somaschi}, \citenamefont {De~Santis}, \citenamefont {Lema{\^\i}tre} \emph {et~al.}}]{loredo2019generation}%
  \BibitemOpen
  \bibfield  {author} {\bibinfo {author} {\bibfnamefont {J.}~\bibnamefont {Loredo}}, \bibinfo {author} {\bibfnamefont {C.}~\bibnamefont {Ant{\'o}n}}, \bibinfo {author} {\bibfnamefont {B.}~\bibnamefont {Reznychenko}}, \bibinfo {author} {\bibfnamefont {P.}~\bibnamefont {Hilaire}}, \bibinfo {author} {\bibfnamefont {A.}~\bibnamefont {Harouri}}, \bibinfo {author} {\bibfnamefont {C.}~\bibnamefont {Millet}}, \bibinfo {author} {\bibfnamefont {H.}~\bibnamefont {Ollivier}}, \bibinfo {author} {\bibfnamefont {N.}~\bibnamefont {Somaschi}}, \bibinfo {author} {\bibfnamefont {L.}~\bibnamefont {De~Santis}}, \bibinfo {author} {\bibfnamefont {A.}~\bibnamefont {Lema{\^\i}tre}}, \emph {et~al.},\ }\bibfield  {title} {\bibinfo {title} {Generation of non-classical light in a photon-number superposition},\ }\href@noop {} {\bibfield  {journal} {\bibinfo  {journal} {Nature Photonics}\ }\textbf {\bibinfo {volume} {13}},\ \bibinfo {pages} {803} (\bibinfo {year} {2019})}\BibitemShut {NoStop}%
\end{thebibliography}%
%%%%%%%%    Appendix    %%%%%%%%%
%%%%%%%%%%%%%%%%%%%%%%%%%%%%%%%%%
\newpage 
\onecolumngrid
\clearpage

\pagenumbering{arabic}
\appendix
\renewcommand{\thesection}
{\Alph{section}}
\renewcommand{\thefigure}{\thesection\arabic{figure}}
\setcounter{figure}{0} 
\renewcommand{\thetable}{\thesection\arabic{table}}
\setcounter{table}{0}

\section{Numerical simulation of the $\beta$-factor}
\begin{figure*} [h!tbp]
    \centering
    \includegraphics[width=17cm]{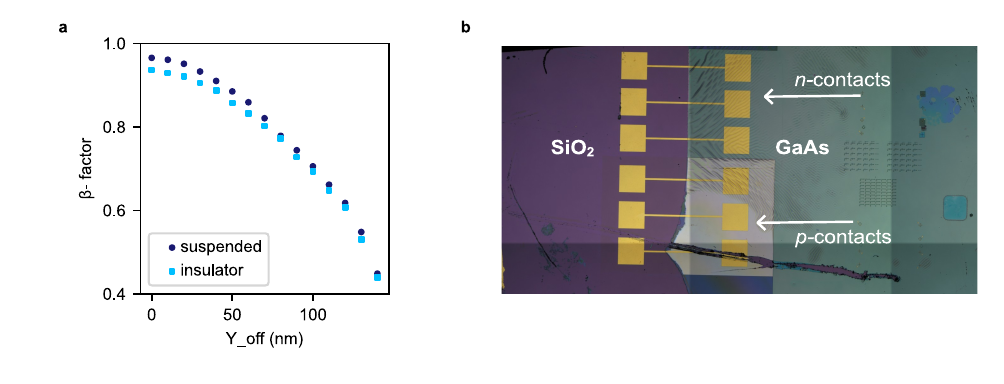}
    \caption{ (a) Comparison of $\beta-$factor for QDs embedded in suspended waveguides and waveguides on insulator. 
    (b) Optical microscope image of the die-to-die bonded chip with fabricated metal contacts and photonic devices.}
    \label{fig:s1}
\end{figure*}

In conventional suspended GaAs waveguides, the high refractive index contrast between GaAs ($n = 3.5$) and vacuum enhances the emitter-mode coupling (or $\beta$-factor) to near-unity enabling deterministic single-photon generation. 
We compute the $\beta$-factor via finite-element modeling of a single-mode nanobeam waveguide with and without an underlying silica layer ($n_\text{SiO$_2$}\sim 1.45$). Figure \ref{fig:s1}a shows the result of the $\beta$ factor calculation as a function of the emitter offset in the waveguide. Despite the lower index contrast offered by silica, at the waveguide center, $\beta$ drops from 96.5\% to 93.6\%. It is expected that adding more Purcell enhancement into a waveguide mode (e.g. via photonic crystal waveguides), $\beta$ can be preserved close to unity. 

\section{Die-to-die bonding and fabrication process}
The fabrication process starts with die-to-die bonding of a silicon chip (around 12$\times$12 mm$^2$) with $~1$ $\mu$m thick thermally-grown silica, and a GaAs chip with embedded QDs (8$\times$8 mm$^2$). We employ an epoxy-based negative tone resist (mr-DWL from Micro Resist Technology GmbH), which is transparent in the infrared and provides a refractive index similar to the one of silica $n\sim1.5$. To provide a strong adhesion between chips, the bonded chips are pressed using a uniform 0.5 kg load and heated on a hot plate at 140 $^\circ$C in vacuum. After the baking process the thickness of the resist is around 260 nm. 

To remove the GaAs substrate we use a two-step wet etching process, first with a solution of nitric acid followed by solution of ammonia peroxide for a selective etching that stops at the AlGaAs sacrificial layer. To remove the sacrificial layer, cold (2 $^\circ$C) hydrochloric acid is used. None of the above wet etching processes attacks the underlying silica. 

Using UV lithography and reactive ion etching (RIE), a via to the \textit{p-}layer of GaAs membrane is opened. Prior to metal contact deposition, an oxygen plasma treatment is performed to remove any native oxide layer from the GaAs membrane and to clear residual resist from the SiO$_2$ surface. The areas designated for metal contacts were then exposed again using UV lithography, followed by the deposition of chromium (10 nm) and gold (200 nm) layers. Chromium serves as an adhesion layer between GaAs and gold. To facilitate reliable wire bonding to the cryostat holder during optical characterization, the metal contacts (or bonding pads) were fabricated onto the silica surface and routed on the membrane as shown in the optical microscope image in Fig.\ref{fig:s1}b. This step was necessary in order to avoid the detachment of the GaAs membrane during ball-bonding, as the adhesion provided by the epoxy layer is not sufficient to withstand the wire-bonding process. 

To fabricate the nanophotonic devices on top of the membrane, 125 keV electron beam lithography is employed using the positive resist ZEP520A. First, shallow-etched grating are defined via RIE and subsequently a deep etching step is used to create the waveguides. The fabricated chip used during the measurements is shown in Fig.~\ref{fig:s1}(b). Both \textit{p-} and \textit{n-}contacts are marked. The scratches on the surface are due to incorrect chip handling but neither affect the quality of the devices nor the electrical contacts. On the right side of the image, the areas with fabricated photonic devices are visible.

\section{Additional characterization of the quantum dot}
\begin{figure}  [h!tbp]
    \centering
    \includegraphics[width=8.5cm]{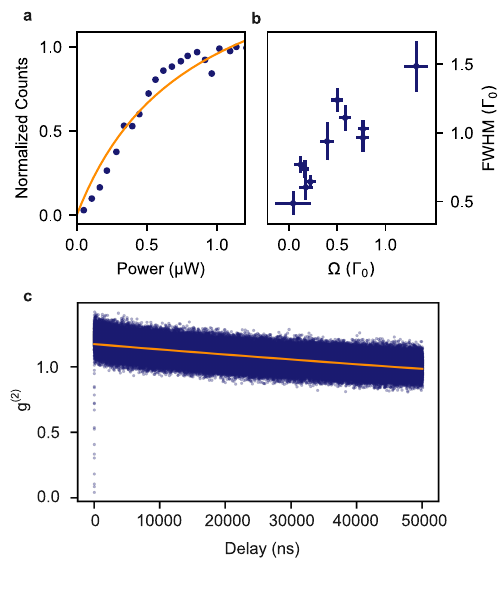}
\caption{Effect of excitation power on intensity and linewidth.
(a) Saturation behavior of the QD's intensity with increasing excitation power. 
(b) Voigt full-width at half-maximum (FWHM) as a function of Rabi frequency. 
(c) Second-order correlation measurement $g^{(2)}(\tau)$ on a long timescale, revealing blinking of the QD emitter: $(17.2 \pm 1.5)\%$.}

    \label{fig:s2}
\end{figure}

The effect of excitation laser power on the emission intensity and linewidth broadening of the QD has been characterized in detail. As shown in Fig.~\ref{fig:s2}(a), the emission intensity increases with power and eventually saturates. This saturation behavior is well described by the model:

\begin{equation}
    I(P) = I_{\infty} \frac{1}{1 + P_{\mathrm{sat}} / P},
\end{equation}

where $I_{\infty}$ is the maximum emission intensity and $P_{\mathrm{sat}}$ is the saturation power. The fitted curve is shown as the orange line. For most experiments, the excitation power was set to 0.7 $\mu$W, just below the saturation threshold, to balance signal strength and spectral purity.

Fig.~\ref{fig:s2}(b) displays the excitation-power-dependent broadening of the QD emission line, plotted as a function of the Rabi frequency. Based on the radiative decay rate extracted from Fig.~\ref{fig:2}(c), the Rabi frequency $\Omega(P)$ at a given power $P$ is given by:

\begin{equation}
    \Omega(P) = \frac{\gamma}{\sqrt{2}} \sqrt{\frac{P}{P_{\mathrm{sat}}}},
\end{equation}

where $\gamma$ is the radiative decay rate. To quantify the power-induced broadening, each emission spectrum was fitted using a Voigt model.
As expected, the linewidth broadening follows the trend of saturation behavior. 

Fig.~\ref{fig:s2}(c) shows the second-order correlation function $g^{(2)}(\tau)$ measured over an extended timescale, revealing blinking behavior of the QD emitter with a magnitude of $(17.2 \pm 1.5)\%$. This level of blinking suggests moderate charge instability in the local environment of the QD. To further mitigate blinking effects, improvements such as minimizing fabrication imperfections, incorporating additional surface passivation, and embedding the emitter in an optical cavity to reduce noise could be explored \cite{davancco2014multiple, liu2018single}. \\

\begin{figure*}[h!tbp]
   \centering
   \includegraphics{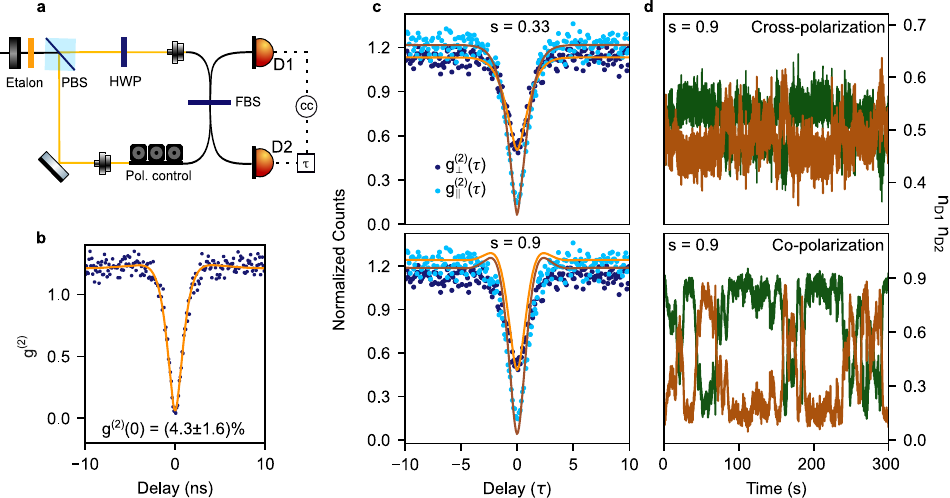}
   \caption{\textbf{Single-photon performance characterization under CW excitation.}
   (a) Schematic of the Mach-Zehnder Interferometer (MZI) setup. Single photons are split into two arms using a half-wave plate (HWP) and a polarizing beam splitter (PBS). Polarization in each arm is controlled by a second HWP and a polarization paddle. The arms are recombined at a 50:50 fiber beam splitter (FBS) and detected by superconducting nanowire single-photon detectors (SNSPDs), D1 and D2. 
   (b) Second-order correlation measurement $g^{(2)}(\tau)$ under sub-saturation excitation ($s = 0.33$), showing strong suppression at zero delay.
   (c) Hong-Ou-Mandel (HOM) interference under CW excitation at two power levels ($s = 0.33$ and $s = 0.9$) in both co- and cross-polarized configurations.
   (d) Normalized single-photon count rates on detectors D1 (green) and D2 (orange) at $s = 0.9$ for different polarization settings, showing interference fringes under co-polarization. Here, $s$ denotes the excitation power normalized to the saturation power.
   }
   \label{fig:suppl3}
\end{figure*}

The single-photon performance of the quantum dot (QD) emitter was investigated under continuous-wave (CW) resonant excitation using a Mach-Zehnder interferometer (MZI), as shown in Fig.~\ref{fig:suppl3}a. Photons were spectrally filtered using a 3~GHz etalon to suppress phonon sidebands prior to entering the interferometer. The beam was equally split into two arms via a polarization beam splitter (PBS) and recombined with a 50:50 fiber beam splitter (FBS). Polarization in each arm was carefully tuned using a half-wave plate (HWP) and a polarization paddle to enable interference.

The single-photon purity was first assessed by measuring the second-order correlation function $g^{(2)}(\tau)$ with one arm of the interferometer blocked. As shown in Fig.~\ref{fig:suppl3}b, the measurement at normalized excitation power $s = 0.33$ revealed a strong anti-bunching signature, with $g^{(2)}(0) = 3.7 \pm 1.9\%$. This confirms highly pure single-photon emission, with negligible multi-photon events.

To evaluate photon indistinguishability, Hong-Ou-Mandel (HOM) interference was performed under CW excitation at $s = 0.33$ and near saturation at $s = 0.9$. The second-order correlations in co-polarized ($g^{(2)}_{\parallel}(\tau)$) and cross-polarized ($g^{(2)}_{\perp}(\tau)$) configurations are presented in Fig.~\ref{fig:suppl3}c. A clear suppression of the central peak in the co-polarized case indicates high indistinguishability of the emitted photons, maintained across a range of excitation powers. The fits (solid lines) to the experimental data (dots) account for decay and dephasing processes to extract the photon overlap fidelity \cite{schofield2022photon}.

Further confirmation of photon coherence was obtained by monitoring the normalized count rates at detectors D1 and D2 for different polarization settings at $s = 0.9$, shown in Fig.~\ref{fig:suppl3}d. In the co-polarized configuration, the two detector outputs exhibited clear intensity oscillations due to interference, whereas the cross-polarized configuration showed minimal modulation, as expected. These oscillations arise from the phase difference $\theta$ introduced between the two interferometer arms, and their contrast scales with the photon indistinguishability and single-photon character\cite{loredo2019generation}.

Overall, these results confirm that the hybrid platform supports highly coherent and indistinguishable single-photon emission under CW excitation, with purity exceeding 96\% and indistinguishability above 91\%, comparable to leading QD sources on pure GaAs platforms.

\section{Linewidth measurements of other quantum dots}

\begin{figure}[h!tbp]
    \centering
    \includegraphics[width=13cm]{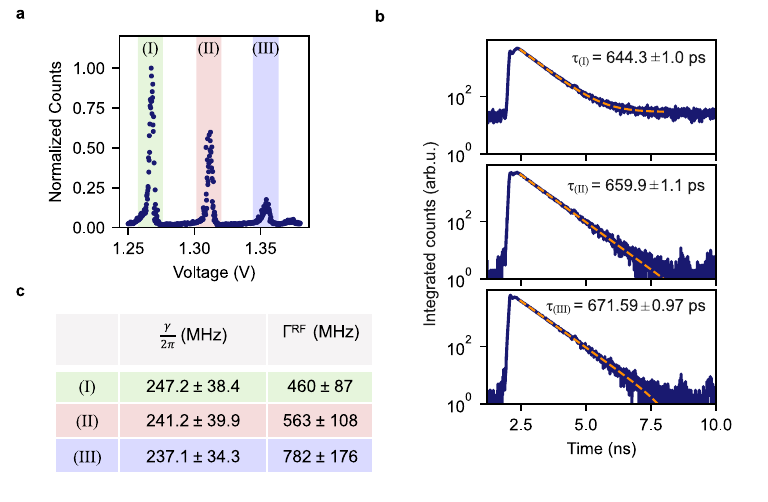}
    \caption[
        \textbf{Linewidth characterization for multiple dipoles.}
    ]{%
        \textbf{Linewidth characterization for multiple dipoles.}
        (a) Emission spectrum as a function of applied voltage, where three distinct lines are visible (marked as (I), (II), and (III), respectively). These lines are likely associated with different charge states of a single QD or potentially originate from neighboring quantum emitters.%
        (b) Time-resolved measurement for each emission line. %
        (c) Comparison of the natural linewidths of the emissions determined from time-resolved measurements, with the linewidth directly registered from the spectrum scanning.%
    }
    \label{fig:suppl4}
\end{figure}

Multiple dipoles are visible in the spectrum, which could correspond to different charge states of a single QD, or alternatively, they may originate from a neighboring quantum emitter. A characterization of the emission linewidths of the three distinct lines observed in the spectrum was performed and is shown in Fig.~\ref{fig:suppl4}(c). The optical setup and excitation frequency of the lasers were adjusted to optimize the emission from the brightest line at 1.27 eV to ensure high signal-to-noise ratio for the primary spectral feature of interest. 

The table presented in Fig.~\ref{fig:suppl4}(c) summarizes the results from Fig.~\ref{fig:suppl4} (a), revealing  the natural linewidth of the emitters, anf d

compares the natural linewidth of the emission lines, determined from time-resolved measurements, with the line broadening values obtained from Lorentzian fits to each emission line. While the lifetimes of the emitters remain consistent within the error margins, the linewidth extracted from spectral fits show visible variation. This suggests that the observed broadening is influenced by factors beyond the intrinsic radiative decay rates, such as environmental noise, charge trapping, power broadening, or instrumental limitations. These results highlight the important role of extrinsic mechanisms in shaping the emission properties, even when emitter lifetimes remain relatively stable. As discussed in the main text, reducing the excitation power and embedding the emitter in an optical cavity could help suppress line broadening and improve spectral stability.

\section{Heterogeneous integration to SiN chips}

\begin{figure}[h!tbp]
    \includegraphics{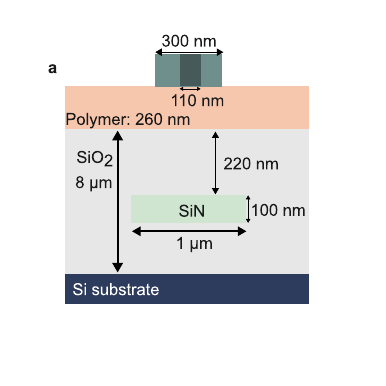}
    \caption{Cross-sectional front view of the coupler of the heterogeneously integrated device.
    }
    \label{fig:suppl5}
\end{figure}

The details and device geometry of the adiabatic taper section are shown in figure \ref{fig:suppl5}. 
The Si$_3$N$_4$ waveguide is embedded in a SiO$_2$ layer with a thickness of 8 $\mu$m and is separated from the chip by a cladding thickness of 220 nm after chemical-mechanical polishing. An additional 260 nm adhesive layer results in a total distance between the GaAs and the SiN waveguide of ~480 nm. This value is sufficiently large to enable low-loss propagation of the SiN waveguide while still allowing for power transfer via mode conversion to GaAs. The mode adapter consists of two linear tapers, the first reducing the GaAs width from 300 nm to 170 nm and the second from 170 nm to a tip that no longer carries the optical mode over a distance of 100 $\mu$m.  With electron beam lithography, the tip width can be as narrow as 60 nm. However, it is found consistently that tips around 100 nm width are sufficient to fully transfer the mode without introducing substantial back-reflection. 
Proximity correction software (Beamfox Technologies ApS) for electron beam lithography is used to pattern the waveguides with better than 10 nm accuracy, which is crucial for achieving high-fidelity tip size over long taper structures. 

\subsection{Transmission model}
The transmission curves in figure \ref{fig:4}d are fitted to a model that includes the grating response and the reflections from the substrate. 
To model the transmission of the grating coupler $T_\text{g}$ we use a Gaussian profile:
\begin{equation}
T_\text{g}(\lambda) = \exp\left(\frac{-(\lambda-\lambda_0)^2}{2\sigma_\lambda^2}\right),
\end{equation}
whereas for the Fabry-P\`{e}rot response $T_\text{FP}$ we use a single-reflectivity $R$ Airy model:
\begin{equation}
T_\text{FP}(\lambda) = \frac{(1-R)^2}{\left((1-R)^2 + 4R\sin^2(2\pi L_\text{opt}/\lambda +\phi_0)\right)},
\end{equation}
where $L_\text{opt}$ is the optical length of the cavity, i.e. approximately $L_\text{opt} = n_{\text{SiO}_2} h$. 
The results of the fit $A\cdot T_\text{g}\cdot T_\text{FP}$ (only $\phi_0$ is set manually for each fit) are given in the table below:
\begin{table}
\caption{Results of fits in figure \ref{fig:4}d}
\begin{center}
\begin{tabular}{|l|c|c|c|c|c|}
\hline
 Device & $\lambda_0$ [nm] & $\sigma_\lambda$ [nm] & $R$ & $L_\text{opt}$ [$\mu$m] & $A\cdot10^3$ [a.u]\\ 
 \hline
 Reference & $(929\pm 1)$ & $(17\pm 2)$ & $(0.26\pm 0.01)$ & 12.5 $\mu$m & $(2.9 \pm 0.1) $ \\  
  Devices & $(930\pm 1)$ & $(21\pm 2)$ & $(0.27\pm 0.03)$ & 11.9 $\mu$m & $(1.22 \pm 0.13) $\\
\hline
\end{tabular}
\end{center}
\end{table}
The results are consistent from reference to devices, and allows for an estimate of the insertion loss of two tapers as $(-3.7\pm 0.3)$ dB, or an efficiency per-taper of $(0.65 \pm 0.05)$. We note that since each taper is $100$ $\mu$m long, and typical GaAs nanobeam waveguides provide a scattering-induced loss of $\sim 7$ dB/mm, at least -0.7 dB loss per taper is expected in the absence of other fabrication imperfections.

\end{document}